\documentclass[prb, superscriptaddress, showpacs,floatfix, letterpaper, twocolumn]{revtex4} 
\usepackage{hyperref}
\usepackage{amssymb}
\usepackage{amsmath}
\usepackage{float}
\usepackage{graphicx}
\usepackage{epsfig}
\usepackage{epstopdf}
\usepackage[usenames]{color}

\renewcommand*\Im{\text{Im}}
\newcommand*\Tr{\text{Tr}}
\newcommand{\Or}{\mathord{\mathcal{O}}}
\newcommand*\bs{\boldsymbol}
\newcommand*\rmi{{\rm i}}
\newcommand*\rme{{\rm e}}
\newcommand*\rmd{{\rm d}}

\begin{document}

\author{Sebastian Fuchs} 
\affiliation{Institut f\"{u}r theoretische Physik, Georg-August-Universit\"{a}t G\"{o}ttingen, 37077 G\"{o}ttingen, Germany}

\author{Emanuel Gull} 
\affiliation{Department of Physics, Columbia University, New York, NY 10027, USA} 

\author{Matthias Troyer}
\affiliation{Theoretische Physik, ETH Z\"urich, 8093 Z\"urich, Switzerland}

\author{Mark Jarrell}
\affiliation{Louisiana State University, Baton Rouge, LA 70803, USA}

\author{Thomas Pruschke} 
\affiliation{Institut f\"{u}r theoretische Physik, Georg-August-Universit\"{a}t G\"{o}ttingen, 37077 G\"{o}ttingen, Germany}

\title{Spectral properties of the three-dimensional Hubbard model}

\date{\today}

\begin{abstract}
  We present momentum resolved single-particle spectra for the
  three-dimensional Hubbard model for the paramagnetic and
  antiferromagnetically ordered phase obtained within the dynamical
  cluster approximation.  The effective cluster problem is solved by
  continuous-time Quantum Monte Carlo simulations.  The absence of a
  time discretization error and the ability to perform Monte Carlo
  measurements directly in Matsubara frequencies enable us to
  analytically continue the self-energies by maximum entropy, which is
  essential to obtaining momentum resolved spectral functions for the
  N\'eel state. We investigate the dependence on temperature and
  interaction strength and the effect of magnetic frustration
  introduced by a next-nearest-neighbor hopping. One particular
  question we address here is the influence of the frustrating
  interaction on the metal-insulator transition of the
  three-dimensional Hubbard model.
\end{abstract}


\maketitle

\section{Introduction}

One of the paradigms for correlation effects and competing orders in
solid state physics is the Hubbard model
\cite{hubbard,gutzwiller,kanamori}
\begin{align}
  \label{eq:hamiltonian}
  H =& 
  -t \sum\limits_{\langle i,j\rangle \sigma} 
  c^{\dagger}_{i\sigma} c^{\phantom\dagger}_{j\sigma}
  -t' \sum\limits_{\langle\langle i,j\rangle\rangle \sigma}
  c^{\dagger}_{i\sigma} c^{\phantom\dagger}_{j\sigma} \nonumber\\
  &  + U \sum\limits_{i} \left(n_{i\uparrow}-\frac{1}{2}\right)
  \left(n_{i\downarrow}-\frac{1}{2}\right) \;.
\end{align}
The operators $c^{\dagger}_{i\sigma}$ ($c_{i\sigma}$) create
(annihilate) an electron with spin $\sigma \in \{\uparrow,
\downarrow\}$ at lattice site $i$, $n_{i\sigma} =
c^{\dagger}_{i\sigma}c_{i\sigma}$ is the particle number operator, $t$
describes the hopping between neighboring sites (denoted by $\langle
i, j \rangle$), $t'$ the hopping between \emph{next}-nearest neighbors
(denoted by $\langle\langle i, j \rangle\rangle$) and $U$ implements
the local Coulomb repulsion.

The Hubbard model, despite its simple structure, can only be
solved exactly in one \cite{1d_hm} and infinite \cite{dmftrev} spatial
dimensions. There are several expectations one can deduce from general
energetic arguments and in particular from the connection between the
Hubbard model and the Heisenberg hamiltonian in the limit of large
interaction strength $t/U\to0$.\cite{fulde_book} For a
three-dimensional simple-cubic lattice and $t'=0$, the Hubbard model
at half filling shows antiferromagnetic order at finite temperature
for any value $U>0$.  The doubling of the unit cell causes this
ordered state to be an insulator. With increased next-nearest-neighbor
hopping this transition is suppressed and a Mott-Hubbard
metal-insulator transition (MH-MIT) is expected to appear in the
paramagnetic state at some non-zero critical value $U_c$ of the
interaction.

Quantum Monte Carlo (QMC) methods are powerful tools that enable the
controlled calculation of properties of large interacting quantum
many-particle systems. Examples include spin models and many bosonic
systems.  However, simulations of fermionic models away from
particle-hole symmetry are often severely hampered by the fermionic
sign problem.\cite{sign} In particular, the identification of ordered
phases, which requires a reliable finite-size scaling, becomes
exceedingly complicated. Independent of the sign problem, the
\emph{direct} investigation of the properties of ordered phases
possibly present in the thermodynamic limit is not possible, since any QMC
simulation is performed on a finite system which cannot exhibit a
spontaneously broken symmetry.

Therefore an approximation scheme allowing (i) calculations in the
thermodynamic limit while (ii) including dynamical correlations in a
controlled way is highly desirable. The dynamical mean-field theory
(DMFT) \cite{dmftrev, dmftrev2} and its cluster extensions
\cite{clusterreview} are such theories.  The DMFT maps the lattice
problem onto an effective single-site impurity model, at the cost of
neglecting non-local many-body correlation effects. In many cases, the
sign problem of the resulting impurity model is either absent or
manageable.  However, these non-local correlation effects are often
crucial for the interplay between Fermi liquid and more exotic states
of matter. Cluster mean-field theories are extensions of the DMFT to
finite clusters, re-introducing non-local (short-ranged) correlations
in a systematic manner but at the same time increasing the complexity
of a simulation.  In the limit of infinite cluster size they become
exact. The controlled extrapolation of cluster results to the
thermodynamic limit is often feasible in practice.\cite{fuchs10,submatrix}

DMFT has proven to be a very powerful tool for studying the
fundamental aspects of the MH-MIT \cite{dmftrev2}. Within the DMFT,
however, the MH-MIT is completely hidden inside an antiferromagnetic
phase, which is insulating by symmetry.\cite{zitzler_afm} Introducing
magnetic frustration, e.\,g., by a non-zero $t'$, tends to suppress the
magnetic order and shift the critical interaction strength $U_c$
toward lower values.  Thus the MH-MIT eventually emerges from the
antiferromagnetic phase for large values of $t'$.\cite{peters_frust_hm}

In low-dimensional systems (single site) DMFT is in general not a good
approximation. In particular, for the one-dimensional Hubbard model,
non-local correlations are in fact dominant,\cite{Koch08} leading to a
complete breakdown of Fermi liquid physics and the formation of a
novel low-energy fixed point, the Luttinger liquid.\cite{1d_hm}
Similarly, in two dimensions a strong influence of spin fluctuations
in the Hubbard model is expected, in particular at and close to half
filling. Since the Mermin-Wagner theorem forbids the formation of an
ordered state in two dimensions at finite temperature, the existing
strong magnetic correlations will lead to correspondingly strong
dynamical fluctuations at $T>0$ and will possibly trigger a similar
breakdown of Fermi liquid physics as in the one-dimensional case.
Evidence for this behavior has indeed been observed in various
numerical
simulations.\cite{dagotto_review,clusterreview,Moukouri01,Onoda03,Parcollet04,Hanasaki06,Zhang07,Liebsch08,Gull08,Park08,gull2009,Werner09}

The importance of short-ranged correlations in the three-dimensional
Hubbard model is less clear and less well studied, and detailed
studies of the phase diagram at high temperature have only recently
begun to appear.\cite{deLeo11,fuchs10}  On the one hand, the precise
value of the N\'eel temperature and critical exponents for the
transition into the antiferromagnetic state will be directly
influenced by the presence of spin fluctuations.\cite{kent} Since
the antiferromagnetically ordered phase is an insulator, we expect the
antiferromagnetic spin fluctuations in the paramagnetic phase to
stabilize the MH-MIT, thus shifting the critical $U_c$ toward lower
values.  Adding frustration by, e.\,g., next-nearest-neighbor hopping
$t'$ will further enhance this effect. One may surmise that the MH-MIT
will eventually emerge from the antiferromagnetic phase as in DMFT. To
our knowledge this has not yet been investigated in detail.

In this work we study the Hubbard model for a three-dimensional cubic
lattice using the dynamical cluster approximation (DCA)
\cite{hettler1, Maier05_dwave} cluster dynamical mean-field algorithm
on clusters of size $18$.  We present momentum resolved
single-particle spectra in the paramagnetic and in the
antiferromagnetic phase, and investigate the influence of frustration
effects caused by a next-nearest-neighbor hopping $t'$.  We focus our
investigation on the interplay of frustration and spin fluctuations in
the vicinity of the paramagnetic metal insulator transition.

\section{Method}

We study the Hubbard model in three dimensions within the DCA to
include both the short- to medium-ranged antiferromagnetic
fluctuations and the possibility of actual long-range
antiferromagnetic order. Since the DCA maps the lattice problem onto
an effective periodic cluster coupled to a dynamic bath, numerically
exact quantum Monte-Carlo (QMC) algorithms are ideally suited to solving
this effective model.

Of particular interest in correlated electron systems are dynamical
correlation functions such as single-particle spectra. However, QMC
provides data only on the imaginary time or frequency axis, and the
necessary analytic continuation of these data has proven to be
difficult. The standard tool used to solve this problem is the maximum
entropy method (MEM).\cite{maxentrev}

Previously, the quasi-standard for simulations of fermionic
many-particle systems was the Hirsch-Fye algorithm,\cite{hirschfye}
which uses a discretization of the imaginary time axis.  An
alternative has evolved in recent years by the development of QMC
algorithms in continuous imaginary time.\cite{rubtsov1, rubtsov2,
  ctaux, werner,gullreview} The absence of a time discretization error
and the possibility of Monte Carlo measurements directly in Matsubara
frequencies \cite{rubtsov1} enhance the quality of the data
significantly \cite{qmc_comparison} and hence enable us to directly
analytically continue self-energies.\cite{sunny} This avoids the
extraction of the self-energies from already continued Green's
functions by a numerically difficult multi-dimensional root finding
algorithm.\cite{root} In this paper we use an implementation of the
continuous-time interaction expansion (CT-INT) QMC algorithm initially
described by Rubtsov and co-workers \cite{rubtsov1, rubtsov2} which
performs a systematic expansion in the interaction term of the
Hamiltonian.

For a simple cubic lattice in three dimensions the dispersion
including nearest- and next-nearest- neighbor hopping reads
\begin{multline}
  \epsilon_{\bs k} = -2t\sum\limits_{i=1}^{3}\cos(k_i) -4t'\left[
    \cos(k_1)\cos(k_2) \right. \\
  \left. + \cos(k_2)\cos(k_3) + \cos(k_1)\cos(k_3)\right]\;,
\end{multline}
where ${\bs k}=(k_1,k_2,k_3)$ is an element of the first Brillouin
zone of the simple cubic lattice. The full bandwidth
\begin{equation}
  W = 
  \begin{cases}
    12t & \mbox{for }|t'|\leq t/4\;,\\
    8t+16|t'| & \mbox{for }|t'|>t/4\;,\\
  \end{cases}
\end{equation}
of this dispersion is used as the energy scale in this paper.

Let us briefly recall the essential aspects of the DCA.  The central
quantity is the single particle Green's function in imaginary time
$\tau$, defined by
\begin{equation}
  G_{\sigma ij}(\tau) = -\langle
  \mathcal{T} c_{\sigma i}(\tau) c^{\dagger}_{\sigma j} \rangle\ .
\end{equation}
Here $\mathcal{T}$ is the imaginary-time ordering operator,
$\langle\cdot\rangle$ denotes a thermal expectation value, and
$c_{\sigma i}(\tau) = e^{-H\tau} c_{\sigma i} e^{H\tau}$.  The spatial
and temporal Fourier transform of Green's function is
\begin{align}
  G_{\sigma\bs k}(\rmi\omega_n) &= \frac{1}{N}\sum\limits_{ij}
  \exp\left[\rmi{\bs k}({\bs R}_i-{\bs R}_j)\right] \times \nonumber \\
  &\times\int\limits_0^{\beta} \!\rmd\tau\, \exp\left(\rmi\omega_n\tau\right) 
  G_{\sigma ij}(\tau)\;,
  \label{eq:ft}
\end{align}
where $\bs k$ is located in the first Brillouin zone and
$\omega_n=(2n+1)\pi/\beta$ with $n\in\mathbb{Z}$ and
$\beta=1/k_{\mathrm{B}}T$ denotes the fermionic Matsubara frequencies.  $T$
is the temperature, and $k_{\mathrm{B}}$ the Boltzmann's
constant. With Dyson's equation, we write Green's
function as
\begin{equation}
G_{\sigma\bs k}(i\omega_n)=\frac{1}{i\omega_n+\mu-\epsilon_{\bs k}-\Sigma_{\sigma\bs k}(i\omega_n)}\;,
\end{equation}
thus introducing the single-particle self-energy $\Sigma_{\sigma\bs
  k}(i\omega_n)$ which contains all the many-body correlation effects
in the system and, in general, is a function of both momentum $\bs k$
and energy $\omega_n$.

\begin{figure}
  \centering
  \includegraphics{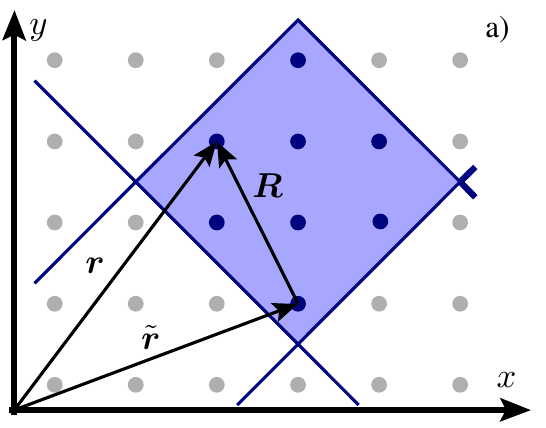}
  \includegraphics{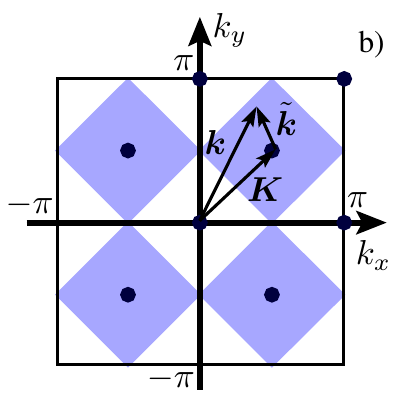}
  \caption{(Color online) Dynamical cluster approximation illustrated for an
    eight-site cluster in two dimensions. In real space (a) the
    origin of a cluster is labeled by $\tilde{\bs r}$. Each site of
    the cluster is identified by $\bs R$. A Fourier transformation
    maps the coordinate ${\bs r} = \tilde{\bs r} + {\bs R}$ of
    each lattice site to a vector $\bs k$ in the first Brillouin zone
    (b). The cluster momentum $\bs K$ now identifies the
    center of a cell in momentum space. All points inside this patch
    can be reached by $\tilde{\bs k}$. The DCA integrates out
    $\tilde{\bs k}$ and thus replaces the full $\bs k$ dependence of
    the lattice by the cells labeled by $\bs K$.}
  \label{fig:dca}
\end{figure}

Within DCA, the full lattice model is approximated by a finite cluster
of size $N$ embedded in a mean field. We tile the first Brillouin zone
into $N$ non-overlapping cells, each represented by its central
momentum $\bs K$ [see Fig.~\ref{fig:dca}(b)
for an example]. The full $\bs k$-dependence of the model is
approximated by the discrete set of $N$ cluster momenta $\bs K$ by
setting $\Sigma_{\sigma\bs k}(\rmi\omega_n)\approx \Sigma_{\sigma\bs
  K}(\rmi\omega_n)$. Averaging over the volume $V$ of the cell
corresponding to cluster momentum $\bs K$, one obtains the quantity
\begin{equation}
  \bar{G}_{\sigma\bs K}(\rmi\omega_n)=\frac{1}{V} \int\!{\rm d}\tilde{{\bs k}}
  \frac{1}{\rmi\omega_n+\mu-\epsilon_{\bs K+\tilde{\bs
        k}}-\Sigma_{\sigma\bs K}(\rmi\omega_n)}\;,
  \label{eq:barG}
\end{equation}
which defines an effective non-interacting cluster via
\begin{equation}
  {\cal G}_{\sigma\bs K}(\rmi\omega_n)^{-1} = 
  \bar{G}_{\sigma\bs K}(\rmi\omega_n)^{-1}+\Sigma_{\sigma\bs K}(\rmi\omega_n)\;.
\end{equation}
With this set of quantities, a suitable method to solve the effective
cluster defined by ${\cal G}_{\sigma\bs K}(\rmi\omega_n)$ and the
interaction $U$, one can determine the self-energy $\Sigma_{\sigma\bs
  K}(\rmi\omega_n)$ iteratively as depicted in Fig.~\ref{fig:loop}.
\begin{figure}
  \centering
  \begin{equation*}
    {\cal G}_{\boldsymbol K}\ \longrightarrow
    \ \text{\setlength\fboxsep{7pt}\fbox{QMC Cluster Solver}}\ 
    \longrightarrow\  G_{\boldsymbol K}
    \vspace*{-3mm}
  \end{equation*}
  \begin{center}
    \begin{minipage}[c]{0.49\linewidth}
      \begin{equation*}
        \uparrow
      \end{equation*}
    \end{minipage}
    \begin{minipage}[c]{0.49\linewidth}
      \begin{equation*}
        \downarrow
      \end{equation*}
    \end{minipage}
  \end{center}
  \vspace*{0mm}
  \begin{equation*}
    ({\cal G}_{\boldsymbol K})^{-1} = (\bar G_{\boldsymbol K})^{-1} +
    \Sigma_{\boldsymbol K}
    \quad\quad\quad
    \Sigma_{\boldsymbol K} = ({\cal G}_{\boldsymbol K})^{-1} - (G_{\boldsymbol K})^{-1}
    \vspace*{-1mm}
  \end{equation*}
  \begin{center}
    \begin{minipage}[c]{0.49\linewidth}
      \begin{equation*}
        \uparrow
      \end{equation*}
    \end{minipage}
    \begin{minipage}[c]{0.49\linewidth}
      \begin{equation*}
        \downarrow
      \end{equation*}
    \end{minipage}
  \end{center}
  \begin{equation*}
    \bar G_{\boldsymbol K} = \frac{1}{V} \int\!{\rm d}\tilde{{\boldsymbol{k}}} \left(
      \mathrm{i}\omega_n + \mu - \epsilon_{{\boldsymbol K+\tilde{\boldsymbol{k}}}} - \Sigma_{\boldsymbol K} \right)^{-1}
  \end{equation*}
  \caption{Self-energy $\Sigma_{\sigma\bs K}(\rmi\omega_n)$ is
    determined self-consistently by iterating the depicted procedure
    until convergence is reached. The bottom line shows the
    calculation of the coarse-grained Green's function $\bar G_{\bs
      K}$ by averaging over the momentum patch centered around $\bs K$
    via integrating $\tilde{\bs k}$ over the volume $V$ of the
    patch. The dependency of Green's functions and the self-energy on
    $\rmi\omega_n$ and $\sigma$ is omitted for simplicity. }
  \label{fig:loop}
\end{figure}

From the QMC algorithm used to solve the effective cluster, we obtain
the cluster Green's function $\bar{G}_{\sigma\bs
  K}(\rmi\omega_n)$. Usually one then uses the maximum entropy method
\cite{maxentrev} to analytically continue this quantity to the real
axis. In order to be able to reverse the coarse-graining, i.\,e.,
calculate $G_{\sigma\bs k}(\omega+\rmi 0^+)$ for all $\bs k$ from the
first Brillouin zone, one needs access to the self-energy
$\Sigma_{\sigma\bs K}(\omega+\rmi 0^+)$ \cite{clusterreview}, which
needs to be obtained by numerical inversion of Eq.~\ref{eq:barG}.
While this is feasible in the paramagnetic phase, the matrix structure
appearing in the antiferromagnetically ordered phase (see section
\ref{sec:afm}) renders this approach impractical.

We follow here an alternative route and analytically continue the
self-energy instead \cite{sunny}, which is related to the cluster Green's function by
\begin{equation}
  \Sigma_{\sigma\bs K}(\rmi\omega_n) = {\cal G}_{\sigma\bs
    K}(\rmi\omega_n)^{-1} - G_{\sigma\bs K}(\rmi\omega_n)^{-1} \;,
\end{equation}
i.\,e., an inversion of $G_{\sigma\bs K}(\rmi\omega_n)$.  This inverse
is calculated directly from the Monte Carlo bins using a jackknife
procedure \cite{jackknife} and therefore incorporates a full error
propagation of the covariance matrix.  The bare Green's function
${\cal G}_{\sigma\bs K}(\rmi\omega_n)$ is viewed here as an input
parameter; and error propagation of errors contained in our estimate of
it, which would require error propagation over subsequent iterations,
is not considered here, such that all errors in ${\cal G}_{\sigma\bs
  K}(\rmi\omega_n)$ are neglected.

\begin{figure}
  \includegraphics{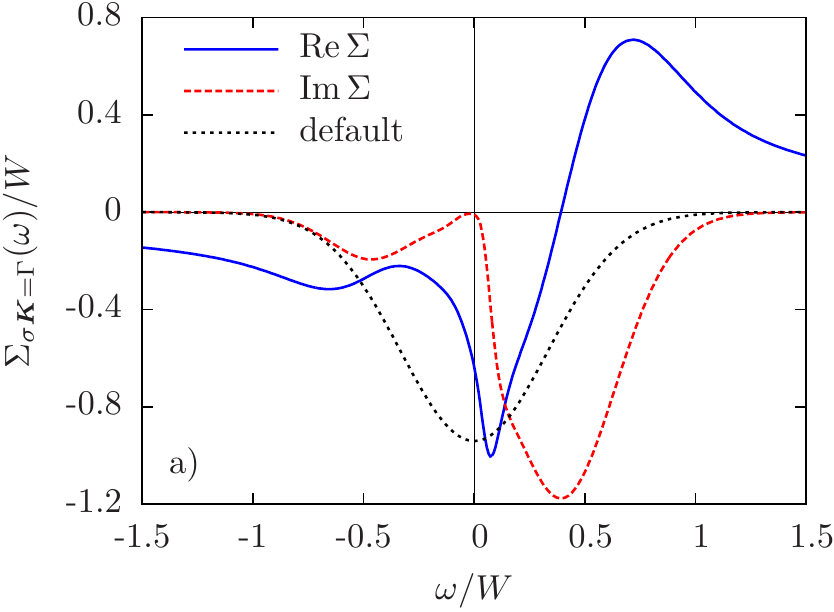}
  \includegraphics{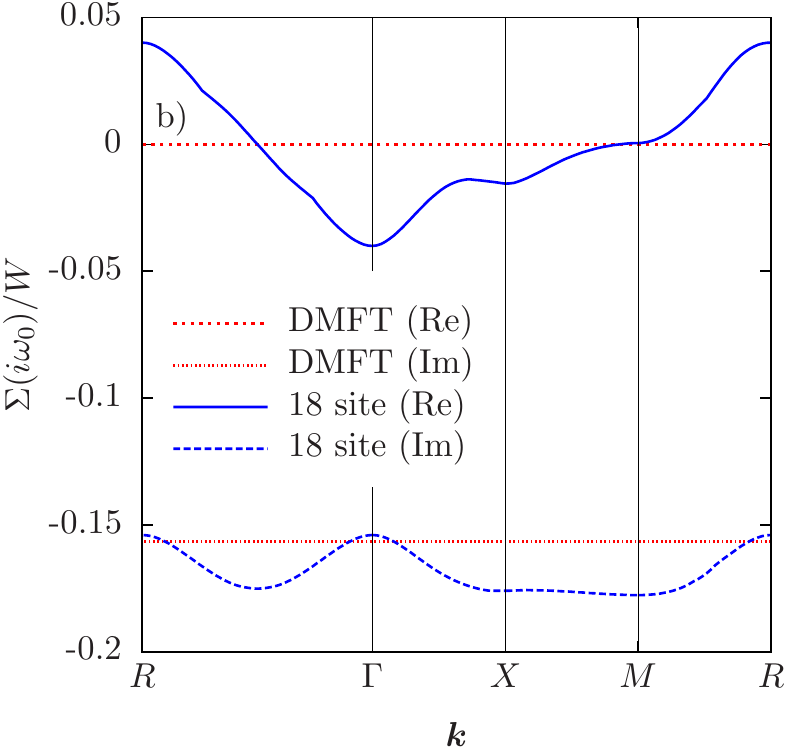}
  \includegraphics{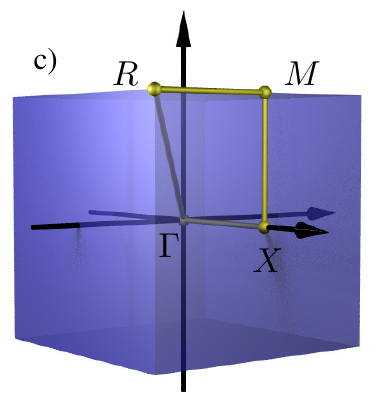}
  \caption{(Color online) (a) Self-energy on the real-frequency axis
    in the paramagnetic phase for $\bs K=\Gamma$, $U=W$, $t'=0$, and
    $T=0.02\,W$ (a). The default model that entered the MEM
    calculation of the imaginary part is also shown. The real part is
    obtained from the imaginary part via Eq.~\ref{eq:kramers}. (b)
    Real and imaginary parts of the interpolated self-energy
    $\Sigma_{\bs k}(\mathrm{i}\omega_0)$ for the smallest Matsubara
    frequency for an 18-site cluster, $U = 0.67\,W$ and $T = 0.03\,W$
    at half filling. The self-energy values at momenta between the
    discrete cluster momenta are obtained from an interpolation using
    Akima splines \cite{akima} in three dimensions. The horizontal
    straight lines denote DMFT results. The interpolation follows the
    path along the high symmetry points of the first Brillouin zone
    depicted in panel (c). For an estimate of finite-size effects in
    (b) see also Ref.~\cite{submatrix}}
  \label{fig:selfpara}
\end{figure}

The analytic continuation of the self-energy from imaginary to real
frequencies is then performed by the maximum entropy method,
\cite{maxentrev} using a standard implementation of the algorithm
following \cite{bryan}. To accurately continue self-energies with MEM,
their high frequency behavior has to be known \cite{sunny}. To this
end we perform a high-frequency expansion of the self-energy
\begin{equation}
  \Sigma_{\sigma\bs K}(\rmi\omega_n) = \Sigma^0_{\sigma} +
  \frac{\Sigma^1_{\sigma}}{\rmi\omega_n} + \Or((\rmi\omega_n)^{-2})\;, 
\end{equation}
where the coefficients are given by (see Appendix\ref{app:tails})
\begin{equation}
  \label{eq:parahigh1}
  \Sigma^0_{\sigma} = U\left( \langle n_{-\sigma} \rangle -
    \frac{1}{2} \right) 
\end{equation}
and
\begin{equation}
  \label{eq:parahigh2}
  \Sigma^1_{\sigma}= U^2 \langle n_{-\sigma}\rangle \left( 1 -
    \langle n_{-\sigma}\rangle \right)\;.
\end{equation}
We now define the quantity
\begin{equation}
  \label{eq:sigmaprime}
  \Sigma'_{\sigma\bs K}(\rmi\omega_n) := \frac{\Sigma_{\sigma\bs
    K}(\rmi\omega_n) - \Sigma_\sigma^0}{\Sigma_\sigma^1}\;.
\end{equation}
Since the average number density $\langle n_{-\sigma} \rangle$ is a
Monte Carlo measurement, we estimate $\Sigma'_{\sigma\bs
  K}(\rmi\omega_n)$ and its covariance matrix by a jackknife
procedure.  The rescaled self-energy $\Sigma'_{\sigma\bs
  K}(\rmi\omega_n)$ as function of Matsubara frequencies is related to
the imaginary part $\Im\, \Sigma'_{\sigma\bs K}(\omega+i0^+)$
on the real frequency axis through the Hilbert transform
\begin{equation}
  \Sigma'_{\sigma\bs K}(\rmi\omega_n) =
  -\frac{1}{\pi} \int\limits_{-\infty}^{\infty}\!\rmd\omega'
  \frac{\Im\, \Sigma'_{\sigma\bs K}(\omega')}{\rmi\omega_n-\omega'}\;.
\end{equation}
By virtue of the rescaling Eq.~\ref{eq:sigmaprime} we furthermore have 
\begin{equation}
  -\frac{1}{\pi}\int\limits_{-\infty}^{\infty}\! \rmd\omega\,
  \Im\, \Sigma'_{\sigma\bs K}(\omega) = 1 \;,
\end{equation}
i.\,e., the spectral function $-\frac{1}{\pi}\Im\,
\Sigma'_{\sigma\bs K}(\omega)$ is non-negative, normalized to unity, and
can thus be calculated by the MEM from the data on the imaginary axis.
The real part of the self-energy then follows from the Kramers-Kronig
relation
\begin{equation}
  \label{eq:kramers}
  \mbox{Re}\,\Sigma_{\sigma\bs K}(\omega) =  -\frac{1}{\pi}
  \,\mbox{P}\!\!\!\int\limits_{-\infty}^{\infty} \!\rmd\omega'
  \frac{\Im\, \Sigma_{\sigma\bs K}(\omega')}{\omega-\omega'} + \Sigma^0_{\sigma}\;,
\end{equation}
where $\mbox{P}\!\int$ denotes a principal value integral. An example
for a full self-energy on the real-frequency axis is shown in
Fig.~\ref{fig:selfpara}(a).
An interpolation of the coarse-grained self-energies yields the
self-energy $\Sigma_{\sigma\bs k}(\omega)$ for all momenta $\bs k$ of
the Brillouin zone. We use a three-dimensional interpolation based on
Akima splines \cite{akima} which provide a smooth interpolation along
the momentum points while avoiding spurious oscillations. Finally,
the single-particle spectral function $A_{\sigma\bs k}(\omega)$ is
calculated using Dyson's equation:
\begin{equation}
  A_{\sigma\bs k}(\omega) = -\frac{1}{\pi}\Im\, \frac{1}{\omega
    + \mu - \epsilon_{\bs k} - \Sigma_{\sigma\bs k}(\omega)}\;.
\end{equation}

\begin{figure*}
  \includegraphics{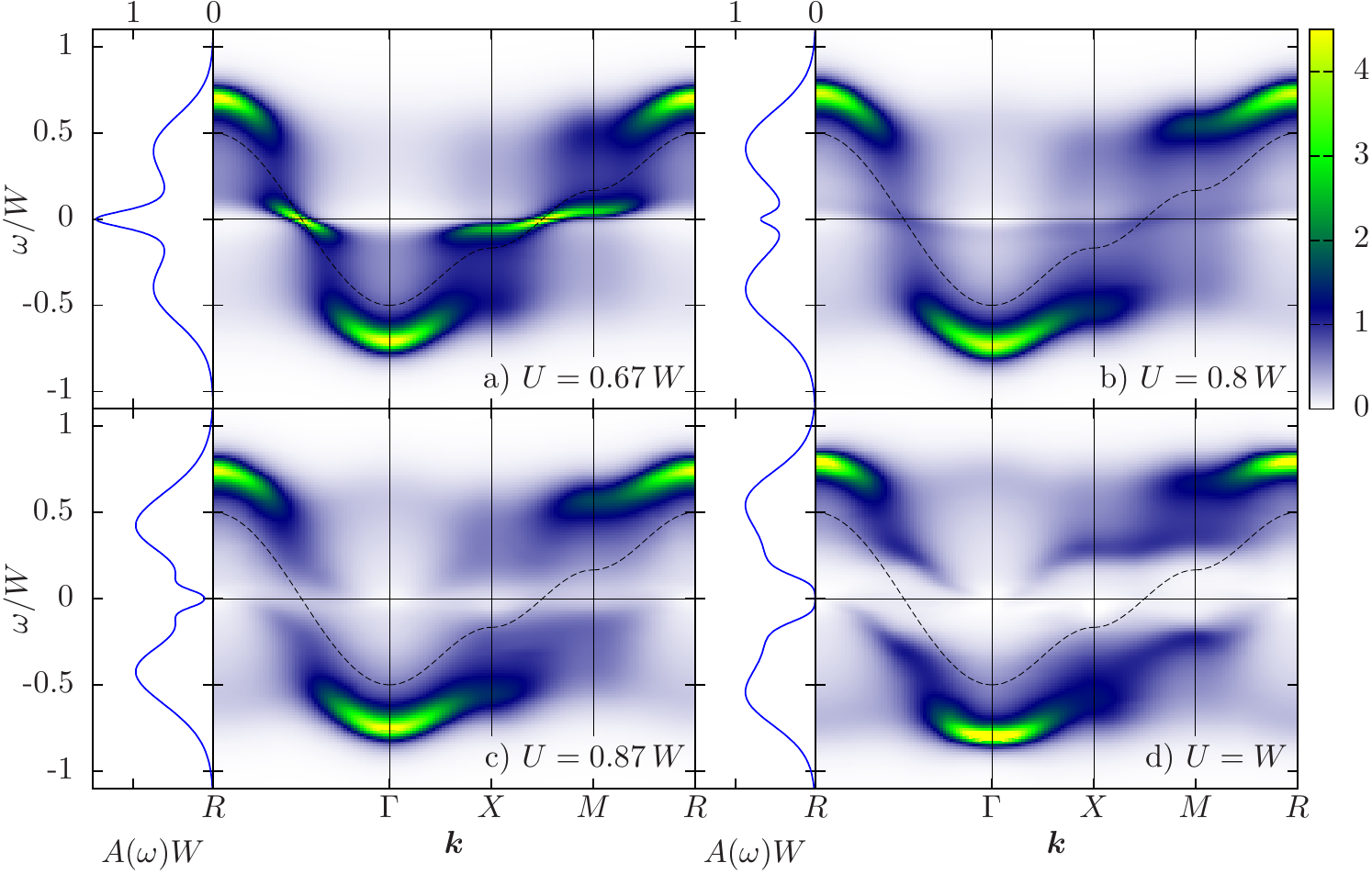}
  \caption{(Color online) Momentum resolved single-particle spectra for
    $T=0.02\,W$ and $t'=0$. The momenta $\bs k$ follow a path along
    high symmetry directions of the first Brillouin zone as depicted
    in Fig.~\ref{fig:selfpara}c. The left part of each diagram shows the
    local single-particle spectrum $A(\omega)$ derived from the direct
    analytic continuation of the Green's function. The dashed line
    denotes the bare dispersion $\epsilon_{\bs k}$.}
  \label{fig:specpara}
\end{figure*} 
The quantity $A_{\sigma{\bs k}}(\omega)$could in principle also be
calculated by an analytic continuation of the Green's function at each
cluster momentum $\bs K$ followed by an interpolation of the Green's
function values directly. However, this procedure approximates the
exactly known momentum dependence of the dispersion $\epsilon_{\bs
  k}$, which causes a significant loss of momentum resolution. We will
not consider it any further.

As an example, Fig.~\ref{fig:selfpara}(b) shows the interpolated
momentum-resolved self-energy for the smallest Matsubara frequency
$\mathrm{i}\omega_0$ (full and dashed lines). The results were
obtained for a cluster with 18 $\bs K$ momenta and a Coulomb repulsion
$U=0.67\,W$ at $T=0.03\,W$. The comparison with the corresponding DMFT
result (dotted lines) shows that the momentum resolution of the DCA
adds significant $\bs k$ dependence to the self-energy. Thus the
many-particle renormalizations acquire a significant $\bs
k$ dependence in DCA, even for the three-dimensional Hubbard model at
comparatively weak coupling.\cite{fuchs10,submatrix}

The available computational resources and the quality of data needed
for high-precision analytic continuation, limit us to study clusters
of comparatively small size. We limit ourselves to a study of a
cluster of size $N=18$ described by the basis vectors ${\bs
  a}_1=(1,1,2)$, ${\bs a}_2=(2,2,-2)$, and ${\bs a}_3=(2,-1,-1)$. This
cluster is the optimal bipartite cluster of this size \cite{kent}
following the criteria proposed by Betts {\it et al.} \cite{betts}.
Since we are primarily interested in identifying trends and basic
physical effects, we do not perform calculations for larger clusters
to obtain a finite-size scaling as would have been necessary, e.g.,
for a precise estimation of the equation of state or the N\'eel
temperature in the thermodynamic limit.\cite{kent,fuchs10}

\section{Properties of the paramagnetic phase}

We begin the discussion of our results by presenting spectral
functions in the paramagnetic phase of the model, i.e., we manually
suppress long-range order.  This allows us to distinguish
dynamical effects coming from fluctuations from effects caused by the
(static) symmetry breaking.

\subsection{Metallic phase}

Figure~\ref{fig:specpara} presents single-particle dispersions in the
paramagnetic phase for $T=0.02\,W$, $t'=0$ and four different Coulomb
repulsions $U=0.67W, U=0.8W, U=0.87W,$ and $U=W$.  The selected $\bs
k$ points follow a path along the high symmetry points of the first
Brillouin zone depicted in Fig.~\ref{fig:selfpara}(c).  Local single
particle spectra are shown on the left of each panel. For small
$U=0.67\,W$ we observe a clear quasi-particle peak at the Fermi level,
both in the density of states (DOS) and the spectral function. The
momentum resolved spectra show that the main contributions to the
quasi-particle peak are situated halfway between the $\Gamma$ and $R$
points and the $X$ and $M$ points, respectively. Comparison of the
peak dispersion in these regions to the non-interacting dispersion
(dashed line) reveals a clear flattening at the Fermi energy, i.\,e.,
an increased effective mass of the quasi-particles.  At higher
energies additional structures---the lower and upper Hubbard
bands---are visible. They follow the curvature of the bare dispersion
but are shifted to higher energies and connected to the quasi-particle
band through broad ``waterfall''-like pieces reminiscent of structures
observed in angle-resolved photoemission spectroscopy of
cuprates.\cite{arpes1} For increasing Coulomb repulsion $U$ the
quasi-particle band at the Fermi energy vanishes and is replaced by an
insulating gap. At the same time the dispersion of the high-energy
structures flattens: the system becomes more localized.  Thus for the
temperature studied a crossover from a metallic dispersion at
$U=0.67\,W$ to a Mott insulating phase at $U=W$ is clearly visible. We
will return to the details of the Mott-Hubbard metal-insulator
transition in section \ref{subsec:MHMIT}.

To make the influence of the momentum dependence of the self-energy in
the spectra more transparent, 
\begin{figure}
  \includegraphics{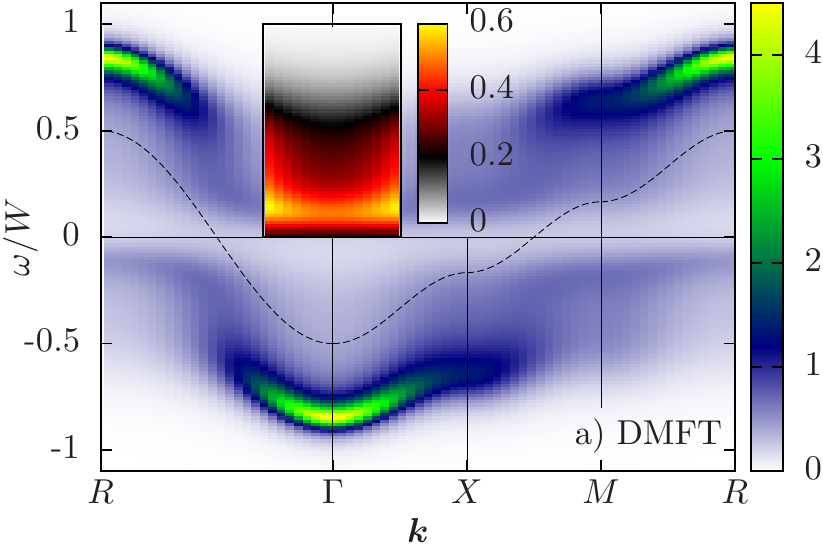}
  \includegraphics{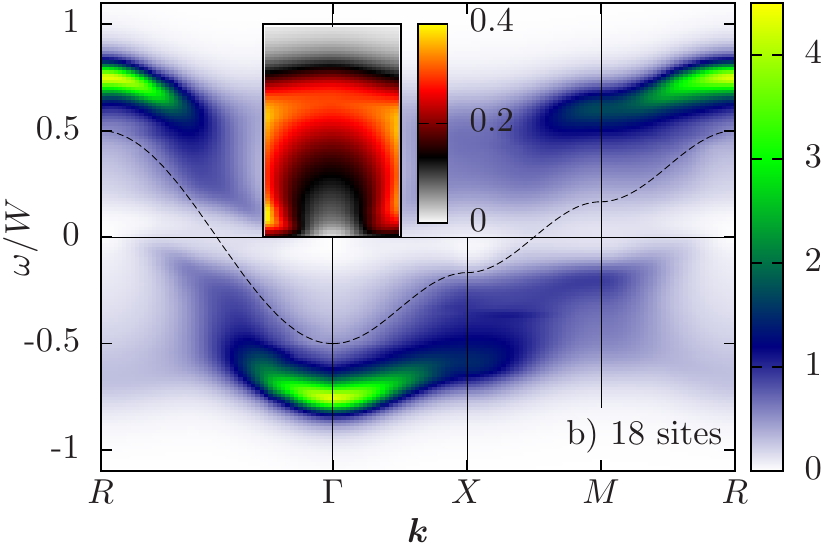}
  \caption{(Color online) Momentum resolved single-particle spectra for
    $T=0.02\,W$, $t'=0$, and $U=0.93\,W$ using a non-dispersive DMFT
    self-energy (a) and a momentum dependent cluster self-energy
    (b). The inset highlights a part of the spectrum using an
    alternative color scheme.}
  \label{fig:dmft}
\end{figure}
Fig.~\ref{fig:dmft} compares an insulating spectrum based on self-energies
of the 18 site cluster to the corresponding spectrum based on a
momentum-independent DMFT self-energy. While many overall features are
similar, there are qualitative differences. For example, a blowup of
the details of the spectra close to the $\Gamma$ point (insets to
Fig.~\ref{fig:dmft}) shows that a substantial part of the DMFT spectrum
around the $\Gamma$ point is located just above the Fermi energy. This
contribution is shifted to higher frequencies in the cluster
calculation, and the curvature is reversed, more resembling the
structure of the lower Hubbard band but with much less spectral
weight. The feature can be interpreted as a precursor of the complete
symmetry with respect to the Fermi energy occurring for spectra in the
antiferromagnetically ordered phase (see section \ref{sec:afm}). Thus,
we attribute these pale reflections of the lower Hubbard band to the
so-called {\it shadow bands} \cite{shadow} arising due to
antiferromagnetic fluctuations not contained in the DMFT simulation.

Next we examine the influence of a next-nearest neighbor hopping $t'$.
In quantum Monte Carlo calculations, a non-zero value of $t'$
introduces a fermionic sign problem even at half filling that may lead
to a significantly larger computational cost.  However, for the
temperatures, Coulomb repulsions and cluster size studied here the
average sign is always greater than 0.94 and thus affects the
efficiency of the simulations only marginally.

\begin{figure*}
  \includegraphics{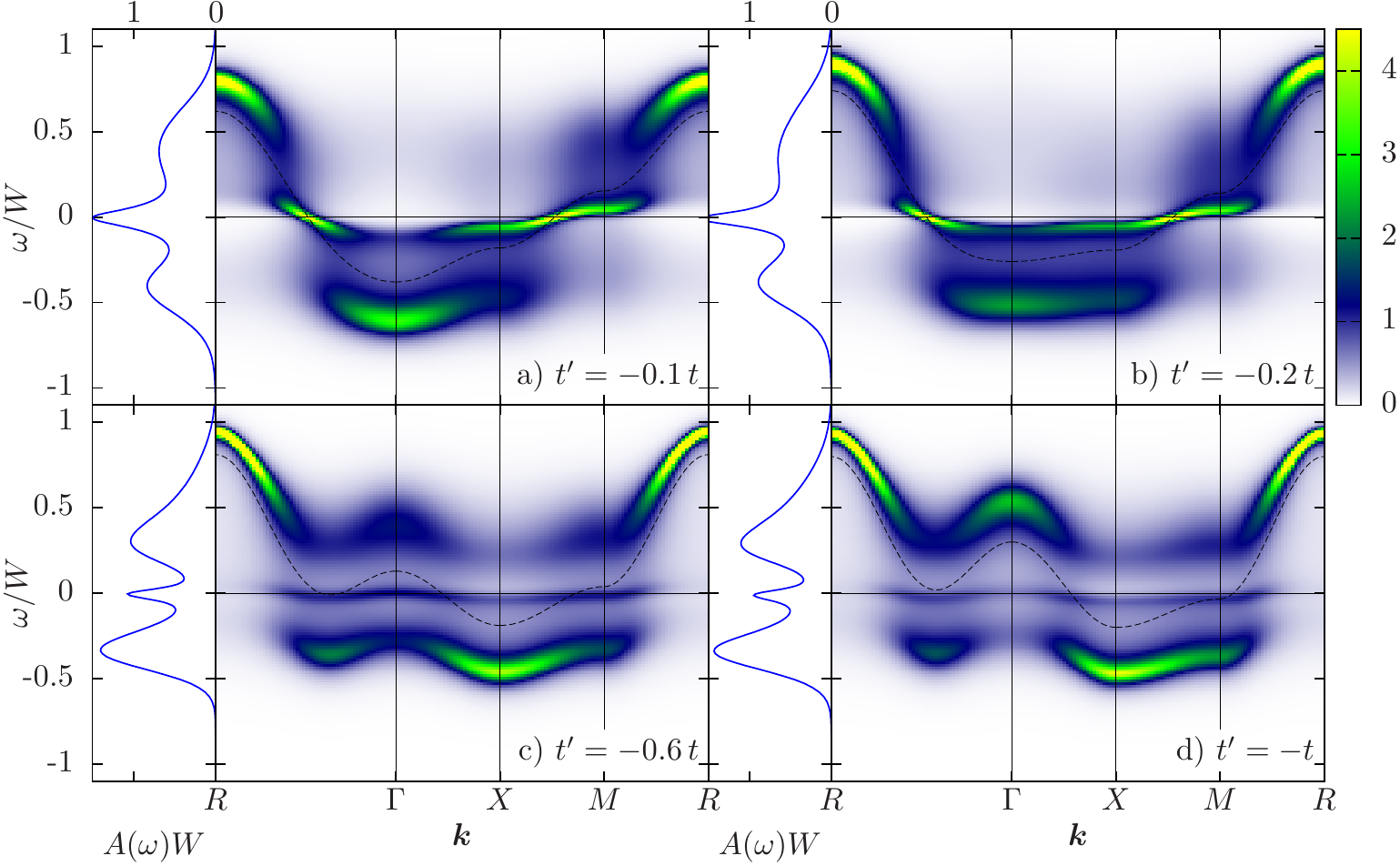}
  \caption{(Color online) Single-particle spectra for $U=0.67\,W$, $T=0.02\,W$ and non-zero $t'$.}
  \label{fig:frustU2m}
\end{figure*}
Fig.~\ref{fig:frustU2m} shows results of calculations for $U=0.67\,W$ and
different values for $t'$. The particle-hole symmetric spectrum of
Fig.~\ref{fig:specpara}(a) becomes more and more asymmetric with increasing
$t'$. We attribute most of these changes to the changes in the bare
dispersion $\epsilon_{\bs k}$. However, while for small to moderate
$t'$ the quasi-particle properties do not seem to change much, one
observes a significant reduction in the spectral weight at the Fermi
energy for larger $t'$, resulting in a reduction of the quasiparticle
peak in the analytically continued spectra.  For example, at $t'=-t$
the integrated weight of the quasi-particle peak is reduced by 50\%
compared to the value for $t'=0$ in Fig.~\ref{fig:specpara}(a). These
observations indicate a reduction of the quasi-particle mass with
increasing $t'$, in accordance with previous DMFT findings
\cite{peters_frust_hm}.

\begin{figure*}
  \includegraphics{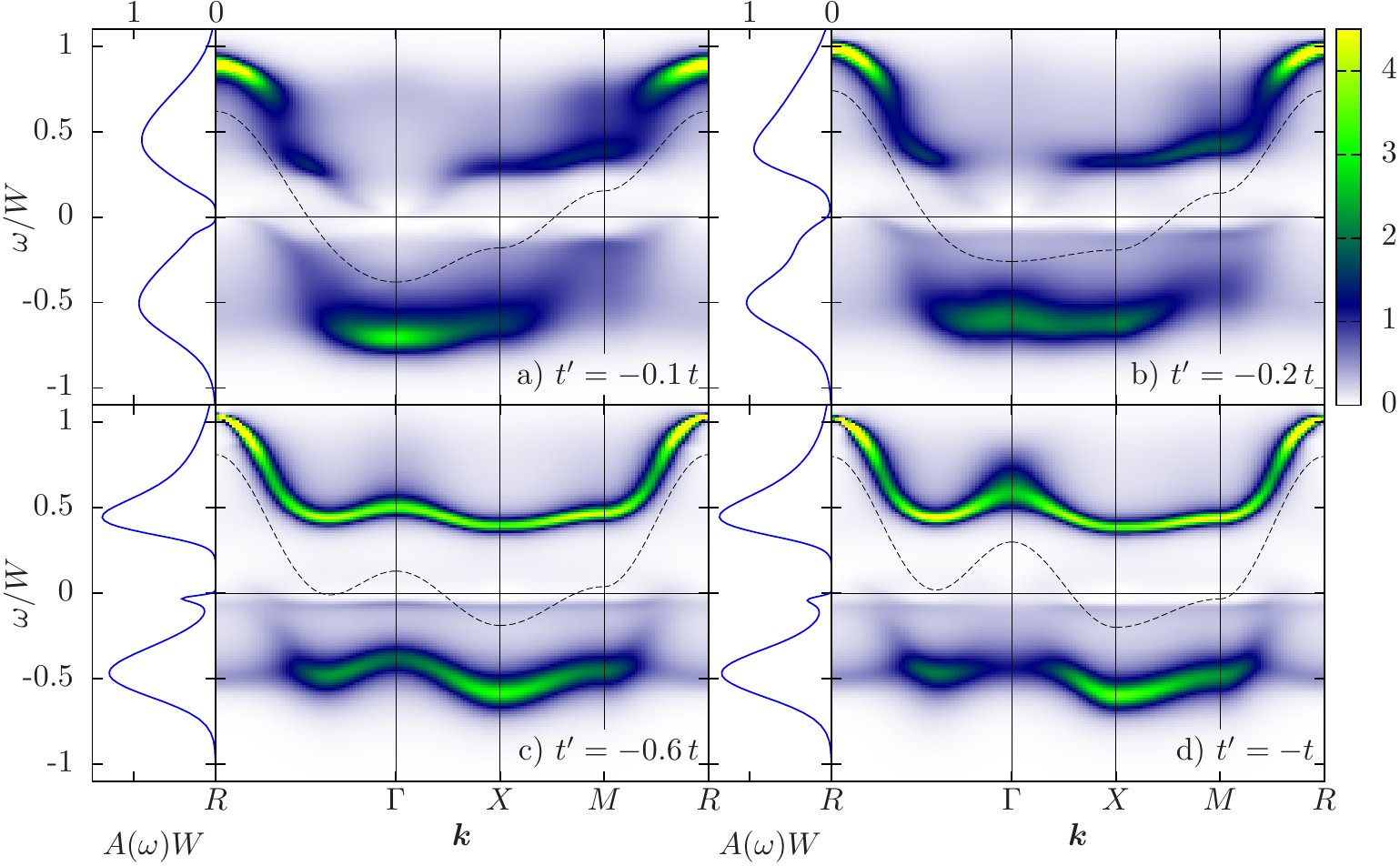}
  \caption{(Color online) Single-particle spectra for $U=W$, $T=0.02\,W$ and non-zero $t'$.}
  \label{fig:frustU3m}
\end{figure*}

Frustration effects on the insulating spectrum in Fig.~\ref{fig:specpara}(d)
($U=W$) are illustrated in Fig.~\ref{fig:frustU3m}.
As for the metallic spectrum, the features present in
Fig.~\ref{fig:specpara}(d) for $t'=0$ initially change only weakly and
in particular the ``shadow'' structures seem to be present, albeit
with reduced weight.  For strong frustration, Hubbard bands become
dominant. These Hubbard bands have a rather well-defined structure and
dispersion reminiscent of the bare dispersion. Furthermore, a peak
develops just below the Fermi energy, which becomes more pronounced
with increasing $t'$ and appears to have only small momentum
dependence.

\subsection{Mott-Hubbard Metal-Insulator Transition}
\label{subsec:MHMIT}

One of the interesting properties of the Hubbard model is the
formation of a correlation driven metal-insulator transition in the
paramagnetic phase, the so-called Mott-Hubbard metal-insulator
transition (MH-MIT). Different from the conventional band or Slater
insulators for even electron number, where the insulating behavior is
due to a completely filled band, the MH-MIT occurs in a partially
filled band, which within a simple single-particle picture would thus
be conducting. Such a transition is believed to frequently be present
in transition metal oxides,\cite{imadarev} and has been discussed
both from an experimental and theoretical point of view over the past
50 years. Note that it is also different from the insulating state
originating from an antiferromagnetic order; the latter implies a
broken translational invariance and can hence be purely explained by
band structure effects (see also section \ref{sec:afm}).

On a simple cubic lattice with only nearest-neighbor hopping the
MH-MIT of the Hubbard model at half filling is completely covered by
the antiferromagnetic phase.\cite{Staudt00} Nevertheless, one may
study it within a generalized mean-field theory by suppressing
long-range antiferromagnetic order in the system. This is accomplished
by enforcing translational symmetry on the bath and thereby preventing
any symmetry-breaking.  One of the clearest signs, numerically, for
identifying the MH-MIT is given by the value of the DOS at the Fermi
energy: A non-zero value at $T=0$ is indicative of a metal, a zero
value of an insulator.  Identifying the MIT at $T>0$ without a
detailed analysis of the temperature dependence is more subtle, but
again the DOS at the Fermi level can serve as ``order parameter": Away
from the MH-MIT the DOS varies smoothly as function of temperature.
When approaching the MH-MIT \emph{at half filling} as function of $U$
or $T$, DMFT analyses suggest that the DOS will show a jump
\cite{dmftrev,Moeller} below a critical end point. Furthermore, the
transition is of first order with a nice hysteresis in the critical
region. This behavior has been confirmed for two-dimensional systems
within cluster DMFT on small \cite{Park08} and larger \cite{gull2009}
clusters.  Note that we do not discuss the approach to the MH-MIT as
function of doping. The behavior of the system at this transition is
very different, and has been addressed by a number of groups both in
DMFT and cluster variants on a wide range of
systems.\cite{Murthy:2002,Georges:2003,Werner09,Ferrero09a,Ferrero09b,clustercompare,Lin10,Sordi:2010}

We estimate the density of states at the Fermi level from the
low-frequency behavior of our Matsubara Green's function
$G_{ii}(\rmi\omega_n)$ that is available as direct Monte Carlo
measurement and does not require analytic continuation. This approach
is based on the relation
\begin{equation}
  \beta
  G_K\left(\frac{\beta}{2}\right)=-\frac{\beta}{2}\int\limits_{-\infty}^{\infty}\!\rmd\omega\frac{A_{\bs
      K}(\omega)}{\cosh\frac{\beta}{2}\omega}
  \stackrel{\beta\to\infty}{\longrightarrow}-\pi\,A_{\bs K}(0)
  \label{eq:gtau}
\end{equation}
for the long-time behavior of the imaginary time Green's function and
the fact that the long-time behavior translates into the low-energy
behavior under the Fourier transform.

\begin{figure}
  \centering
  \includegraphics{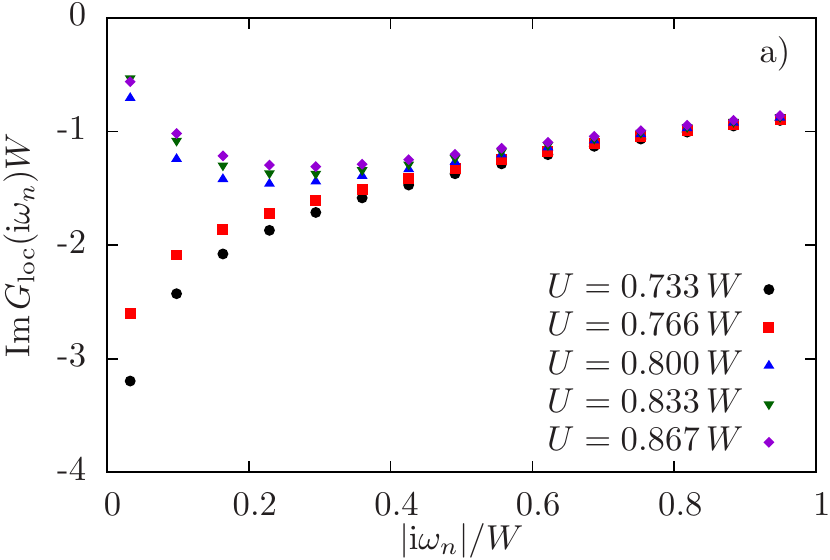}
  \includegraphics{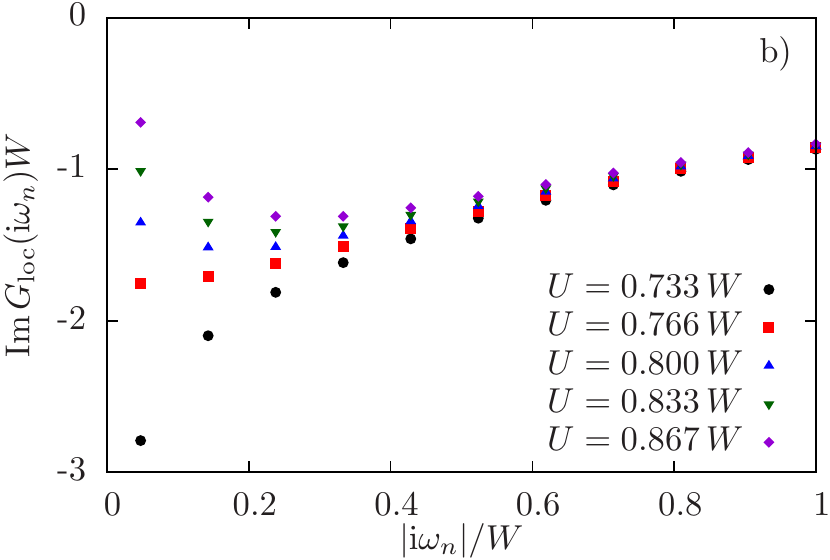}
  \caption{(Color online) Imaginary part of the local Green's
    function in Matsubara frequencies for $t'=0$, $T=0.01\,W$ (a) and
    $T=0.015\,W$ (b). Several values of $U$ around the MH-MIT are
    shown. }
  \label{fig:mit}
\end{figure}

Figure~\ref{fig:mit}(a) 
shows the imaginary part of $G_{\rm loc}(\rmi\omega_n) :=
G_{ii}(\rmi\omega_n)$ for $t'=0$ and $T=0.01\,W$ obtained within the
DCA for a cluster size of 18. The jump from an insulating Green's
function at $U=0.80\,W$ to a metallic solution at $U=0.766\,W$ clearly
shows the location of the MH-MIT at this temperature.  For
$T=0.015\,W$ we could only detect a crossover from insulating to
metallic behavior [see Fig.~\ref{fig:mit}(b)]. This indicates that the
critical endpoint of the MH-MIT transition line is located between
$T=0.01\,W$ and $T=0.015\,W$, substantially below the N\'eel
temperature at this interaction strength [$T_{\rm N}= 0.030(3)\,W$ at
$U=0.8\,W$ \cite{Staudt00, kent}].

Another observable that shows a clear signal of the MH-MIT is the
effective mass, defined as
\begin{equation}
  \label{eq:mass_normal}
  \frac{m^\ast_{\bs k}}{m} = 
  1-\left.
    \frac{\partial Re\Sigma_{\sigma\bs k}(\omega)}{\partial\omega}
  \right|_{\omega=0}\;,
\end{equation}
where $m$ denotes the bare carrier mass. The effective mass at
non-zero (but sufficiently low) temperature may be estimated from the
QMC without resorting to analytical continuations as \cite{mass}
\begin{equation}
 \label{eq:mass}
 \frac{m^*_{\bs k}}{m} \approx 1- \frac{\Im\Sigma_{\sigma\bs k}(\rmi\omega_0)}{\omega_0}\;,
\end{equation}
where $\omega_0$ is the lowest Matsubara frequency (see, e.\,g., Fig.~$10$
in Ref.~\cite{clustercompare} for a comparison between directly
obtained and analytically continued estimates). At the MH-MIT,
$m^*_{\bs k}$ across the Fermi surface exhibits a sharp increase.

\begin{figure}
 \includegraphics{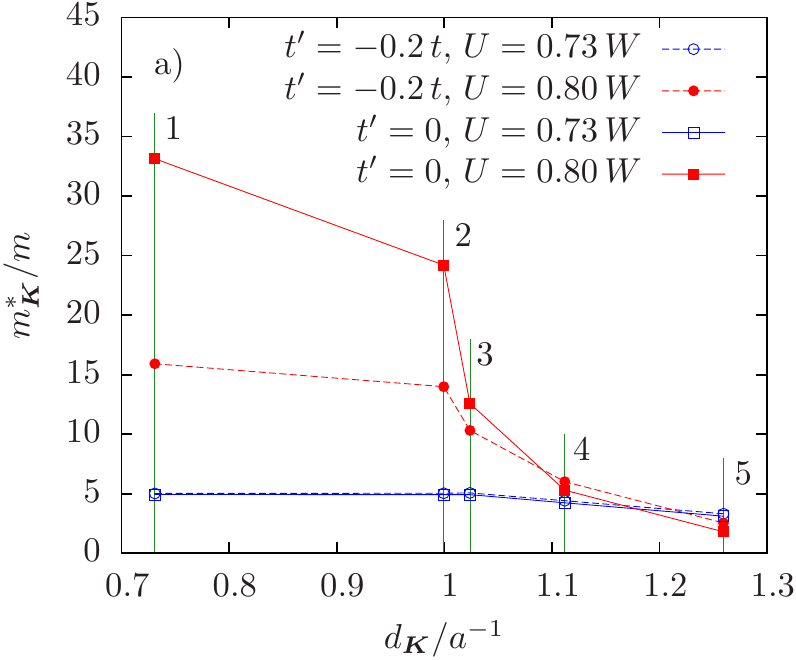}
 \includegraphics[scale=1.1]{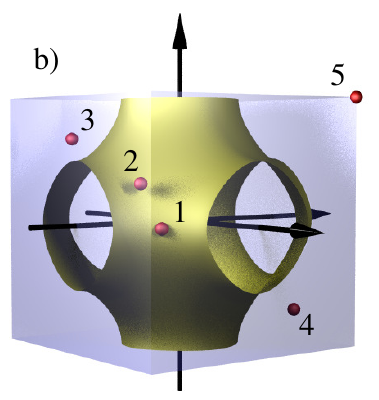}
 \caption{(Color online) (a) Quasi-particle mass $m^*_{\bs K}$ as function of
   mean distance $d$ from the Fermi surface in units of the lattice spacing
   $a$ (a). The distance $d_{\bs K}$ is defined as the average of the
   distances between the Fermi surface and all momenta in the cluster
   cell described by the cluster momentum $\bs K$. The 18 cluster
   momenta reduce to five different masses due to point symmetries. The
   labels 1 to 5 refer to the points depicted in panel (b), which also
   shows the non-interacting Fermi surface for $t'=0$. }
  \label{fig:fermi}
\end{figure}

The estimate for the effective mass obtained that way is shown in
Fig.~\ref{fig:fermi}(a)
for two different values $U=0.73\,W<U_{c}(t'=0)$ and
$U=0.8\,W>U_{c}(t'=0)$ and $t'=0$ and $-0.2\,t$ at $T=0.01\,W$. We do
not show the result for $m^\ast_{\bs k}$ from the interpolated data,
as the division by $\omega_0$ with $\omega_0/W\ll1$ also strongly
enhances spurious artificial oscillations due to interpolation, but
rather we present the masses $m^*_{\bs K}$ for each of the 18 cluster
momenta as functions of their mean distance to the non-interacting
Fermi surface for $t'=0$.  The mean distance is calculated by
averaging the distance to the Fermi surface of all points inside the
cluster cell around $\bs K$. Due to symmetry, some cluster momenta are
equivalent and thus we obtain only five different effective
masses. Figure~\ref{fig:fermi}(b) depicts one representative cluster
momentum for each one of these five equivalence classes.

For both values of $t'$ the ${\bs K}$ dependence of $m^\ast_{\bs K}$
for $U$ deep inside the metallic phase is rather weak, although
nevertheless visible.  The insulating phase, in contrast, shows a
dramatic increase of the effective mass for the cluster momenta near
the Fermi surface. For example, point 1 with the strongest enhancement
is situated on the Fermi surface of the non-interacting system. Thus,
the natural interpretation is that MH-MIT appears first for $\bs
k$ points at or close to the Fermi surface. Points far away from the
Fermi surface, on the other hand, like points 4 and 5 (5, for example,
corresponding to $\Gamma$, respectively $R$), experience only weak
renormalizations. The interpretation is further supported by the
influence of non-zero $t'$, which moderately modifies the mass due to
changing distances of the cluster $\bs K$ points from the Fermi
surface and a different critical $U_c$, but otherwise shows a similar
behavior.  We, however, do not observe a significant variation of
$m^\ast_{\bs K}$ across the Fermi surface. The difference between
points 1 and 2 in Fig.\ \ref{fig:fermi} can be explained by the large
distance of the cluster center from the Fermi surface.

Note that this behavior of the three-dimensional model is different
from the two-dimensional Hubbard model, where the Mott transition
within DCA has been investigated in some
detail.\cite{Werner09,gull2009,clustercompare} In this case, the
so-called ``momentum selective Mott transition'' has been observed,
where different parts of the Fermi surface undergo a metal insulator
transition for different interaction strengths. To conclusively
exclude the possibility of a momentum selective MH-MIT in three
dimensions, larger clusters with a finer grid of $\bs K$ points on
different parts of the Fermi surface would be needed.

\begin{figure}
  \includegraphics{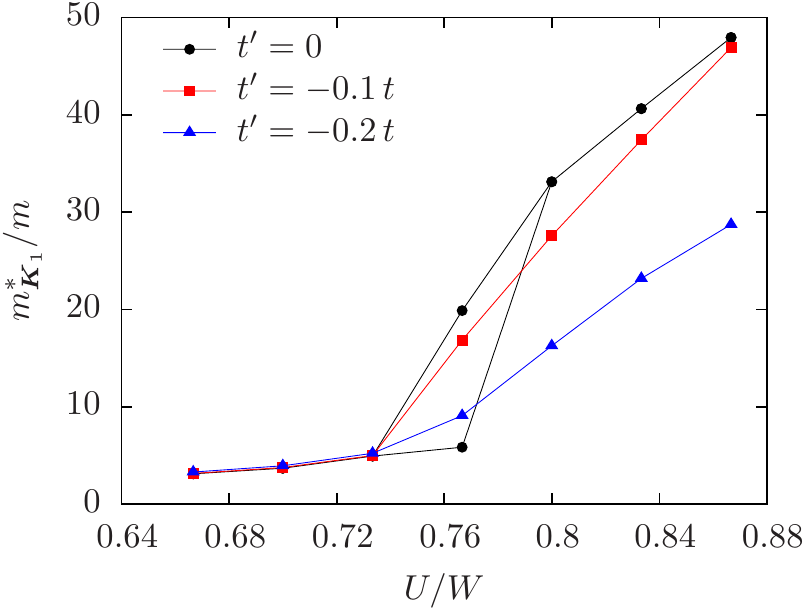}
  \caption{(Color online) Quasi-particle mass estimate $m^*_{\bs
      K_1}$ at the midpoint between $\Gamma$ and $M$ [point 1 in
    Fig.~\ref{fig:fermi}(b)], for $T=0.01\,W$ as function of the
    interaction parameter $U$.}
  \label{fig:mass}
\end{figure}

In the following we focus on cluster momentum 1 in
Fig.~\ref{fig:fermi}(b), which is situated directly on the Fermi
surface, midway between $\Gamma$ and $M$. Since this point exhibits
the strongest mass enhancement in the insulating phase, it is an ideal
candidate for studying the MH-MIT. The effective mass of this $\bs K$
point is plotted in Fig.~\ref{fig:mass}
as a function of $U$.

At $U_{\rm c}=0.766\,W$, both an insulating and a metallic solution
can be stabilized, depending on the initial Green's function used to
start the DCA self-consistency. This behavior indicates a coexistence
region in this regime of interaction parameters and tells us that the
qualitative physical properties of the paramagnetic MH-MIT do not
change at least qualitatively for a true three-dimensional system. The
figure also shows the corresponding curves for next-nearest-neighbor
hopping parameters $t'=-0.1\,t$ and $t'=-0.2\,t$. Here the coexistence
region has vanished at the temperature for which the simulations were
done, while the relatively smooth shape of the curve indicates that
one is still observing a crossover and not yet a sharp phase
transition as in the case of $t'=0$.  This is again in accordance with
previous DMFT calculations, where a reduction of both the critical
temperature and the critical value of $U$ was observed with increasing
$t'$.\cite{peters_mit} It is, however, different from the interaction
transition in the two-dimensional Hubbard model,\cite{gull2009} where
increasing $|t'|$ leads to an increase of the critical interaction
strength.

It would be highly desirable to perform simulations at lower
temperatures for non-zero $t'$, but since the computational effort
necessary increases dramatically with decreasing temperature, we were
not yet able to do these simulations for the time being.

Previous studies of the interaction driven MH-MIT at non-zero
temperature were largely performed on a Bethe lattice in the limit of
infinite dimension within the DMFT approximation,\cite{finite_mit_nrg,
  finite_mit_qmc} respectively, for a two-dimensional Hubbard model
using the correlator projection method \cite{Onoda03,Hanasaki06} or
cluster
DMFT.\cite{Moukouri01,Parcollet04,Zhang07,Liebsch08,Gull08,Park08,gull2009,Werner09}
While the general features of the MH-MIT appear to be rather
insensitive to the actual non-interacting band structure, the details
like critical values for temperature and Coulomb interactions vary
strongly with details of the model as well as the approximations
involved in the computation and finite size effects.

To compare values for lattices with different noninteracting density
of states $\rho(\omega)$, Bulla \cite{bulla} suggested using the
second moment of the DOS
\begin{equation}\label{eq:Weff}
  W_{\mathrm{eff}} = 
  4\sqrt{
    \int\limits_{-W/2}^{W/2} \!\rmd\omega\,\omega^2 \rho(\omega),
  }
\end{equation}
as the characteristic energy scale instead of the bandwidth $W$. From
Eq.~(\ref{eq:Weff}) one obtains $W_{\mathrm{eff}} = W$ for the Bethe
lattice and $W_{\mathrm{eff}} \approx 0.816\,W$ for the simple cubic
lattice, and a rather good agreement of critical values when relating
them to $W_{\mathrm{eff}}$.\cite{bulla,rok} Our result
$U_{\mathrm{c}}=0.77(3)\,W$ for the coexistence region then translates
to $U_{\mathrm{c}} = 0.94(3)\,W_{\mathrm{eff}}$ at
$T=0.012\,W_{\mathrm{eff}}$. For a conventional DMFT calculation,
Refs.~\cite{finite_mit_nrg} and \cite{finite_mit_qmc} located the
coexistence region for this temperature around $U_c =
1.18(2)\,W_{\mathrm{eff}}$. This indicates that for a true
three-dimensional system the critical values of the MH-MIT will be
renormalized, and in particular the critical $U_{\mathrm{c}}$ will be
shifted to lower values.  We attribute these renormalizations to the
short-ranged antiferromagnetic fluctuations present in the DCA. They
will have the tendency to suppress the formation of quasi-particles
and will thus cause the transition to shift to smaller Coulomb
repulsions $U$.  A detailed investigation of the location of the
transition in the thermodynamic limit would require the study of a
range of cluster sizes and a careful finite-size analysis along the
lines of Refs.~\cite{Kent05,Maier05_dwave,fuchs10,submatrix}.

\section{Antiferromagnetic phase}\label{sec:afm}

As previously mentioned the thermodynamically stable low-temperature
phase of the Hubbard model at half filling is, for $t'=0$,
antiferromagnetic. This phase completely covers the
MH-MIT.\cite{Staudt00,peters_frust_hm} Conventionally, the onset of
antiferromagnetic long-ranged order is indicated by a divergence of
the staggered susceptibility upon cooling from the paramagnetic state
at high temperature.\cite{Kent05}  Complementary to a paramagnetic
simulation and analysis of the susceptibility, the symmetry-broken
phase may be simulated directly. While this scheme is less accurate at
determining the location of phase boundaries, it can address the
properties within the ordered phase and thus is relevant for
comparison to experiments within that phase. Furthermore, it often is
desirable to investigate the direct change of quantities in the
presence of competing phases or orders.  For these reasons we present
in this paper results obtained in the antiferromagnetically ordered
state. We will show that within the same framework, with minor
modifications, we can also obtain high-quality spectra from QMC data
in a phase with non-trivial broken symmetry.

\begin{figure}
  \includegraphics{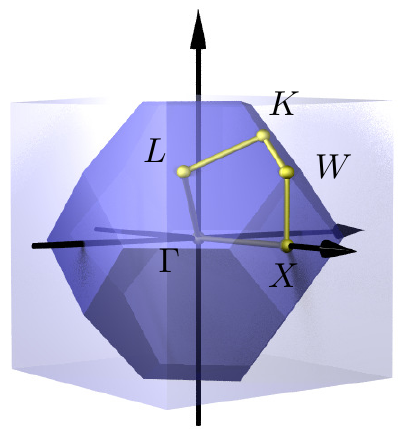}
  \caption{(Color online) Magnetic Brillouin zone of the simple
    cubic lattice and the path along high symmetry points used for the
    presentation of spectra in the antiferromagnetically ordered state
    in Fig.~\ref{fig:anti}}.
  \label{fig:mbz}
\end{figure}

The antiferromagnetic order breaks the translational symmetry of the
lattice, leading to a doubling of the unit cell. This implies that the
first Brillouin zone, correspondingly, halves its size. The resulting
magnetic Brillouin zone (MBZ) is shown in Fig.~\ref{fig:mbz}.

In order to explicitly break the full translational symmetry along
with the SU(2) symmetry, we add a staggered magnetic field $h_{i} =
h_0\,\rme^{\rmi {\bs Q}\cdot{\bs r_i}}$ with ${\bs Q} = (\pi,\pi,\pi)$
to the Hamiltonian of Eq.~(\ref{eq:hamiltonian}) as
\begin{equation}
  \label{eq:hamfield}
  H_{h} = H + \sum\limits_{i} h_{i} m_{i}\;,
\end{equation}
where $m_{i} = n_{i,\uparrow} - n_{i,\downarrow}$ is the spin
polarization at lattice site $i$.  In principle this allows the study
of properties as functions of this staggered field. However, because
such a field is of little experimental relevance, one is
conventionally only interested in the limit $h_0\to0$. If in this
limit a non-zero polarization remains, we have found a state with
spontaneous symmetry breaking.

In the actual simulation we add a small field and explicitly break the
symmetry (in our case we chose $h_0=0.01$) in the initialization of
our iteration process.  The field is switched off after the first few
iterations and the system is allowed to evolve freely. Eventually, the
process converges to a solution with either a vanishing staggered
magnetization $M_i(T)\propto \langle m_i\rangle\,=\,\rme^{\rmi {\bs
    Q}\cdot{{\bs r}_i}}\,m_s(T)=0$, indicating a parameter regime
where the thermodynamically stable state is paramagnetic, or
$M_i\ne0$ and thus an antiferromagnetically ordered state.

To be able to include such a field in our simulations, we
have to ensure that the cluster we use has the proper translational
symmetry with respect to a double unit cell. These clusters are also
referred to as bipartite clusters. We again employ the systematic
classification by Betts \cite{betts} to find the optimal cluster of
this type with 18 sites.  Since the DCA is formulated in momentum
space, the broken translational symmetry introduces explicit
non-diagonal elements in quantities like the Green's function or the
self-energy.  For the following we will adopt the notation
\begin{multline}
  G_{\sigma{\bs K_1},{\bs K_2}}(\rmi\omega_n) = \\ \frac{1}{N}\sum\limits_{ij}
  \exp\left[\rmi\left({\bs K}_1\cdot{\bs R}_i-{\bs K}_2\cdot{\bs R}_j\right)\right] G_{\sigma ij}(\rmi\omega_n)
\end{multline}
as extension of Eq.~(\ref{eq:ft}) for Green's functions with momenta
${\bs K}_1 \neq {\bs K}_2$. With this notation, Green's function in
the antiferromagnetic phase can be represented by the 2$\times$2
matrix
\begin{multline}
  {\bf G}_{\sigma{\bs K}'}(\rmi\omega_n) := 
  \left(
    \begin{array}{ll}
      G^{00}_{\sigma \bs K'}(\rmi\omega_n) & G^{01}_{\sigma \bs K'}(\rmi\omega_n)  \\
      G^{10}_{\sigma \bs K'}(\rmi\omega_n) & G^{11}_{\sigma \bs K'}(\rmi\omega_n)  
    \end{array}
  \right)
  \\
  := 
  \begin{pmatrix}
    G_{\sigma{\bs K'},{\bs K'}}(\rmi\omega_n) & G_{\sigma{\bs K'},{\bs K'}+{\bs Q}}(\rmi\omega_n)  \\
    G_{\sigma{\bs K'},{\bs K'}+{\bs Q}}(\rmi\omega_n) & G_{\sigma{\bs K'}+{\bs Q},{\bs K'}+{\bs Q}}(\rmi\omega_n)  
  \end{pmatrix}
  \label{eq:twobytwo}
\end{multline}
where ${\bs K}'$ is an element of the MBZ. The
symmetry relations $G^{00}_{\sigma\bs K'}(\rmi\omega_n) =
G^{11}_{\bar\sigma\bs K'}(\rmi\omega_n) = -\left(G^{11}_{\sigma\bs
    K'}(\rmi\omega_n)\right)^* = -\left(G^{00}_{\bar\sigma\bs
    K'}(\rmi\omega_n)\right)^*$ and $G^{10}_{\sigma \bs
  K'}(\rmi\omega_n) = G^{01}_{\sigma\bs K'}(\rmi\omega_n) =
G^{10}_{\bar\sigma\bs K'}(\rmi\omega_n) = G^{01}_{\bar\sigma\bs
  K'}(\rmi\omega_n)$ hold for Green's function as well as for the
self-energy. The latter is still defined via Dyson's equation
\begin{equation}
  {\bf \Sigma}_{\sigma\bs K'}(\rmi\omega_n) = {\bs{\cal G}}_{\sigma\bs
    K'}(\rmi\omega_n)^{-1} - {\bf G}_{\sigma\bs
    K'}(\rmi\omega_n)^{-1}\;.
\end{equation}
which, however, now involves quantities which are $2\times2$ matrices.

This matrix structure makes it necessary to adapt the CT-QMC algorithm
accordingly. This can most conveniently be done using a spinor
representation for the field operators, and rewriting the formulas
with these new composite operators. A detailed account of this
procedure will be given elsewhere. Here we just want to note that one
can again perform measurements in Matsubara space directly and thus
obtain an accurate estimate of the self-energy, which can then be
analytically continued as before.

\begin{figure}
  \includegraphics{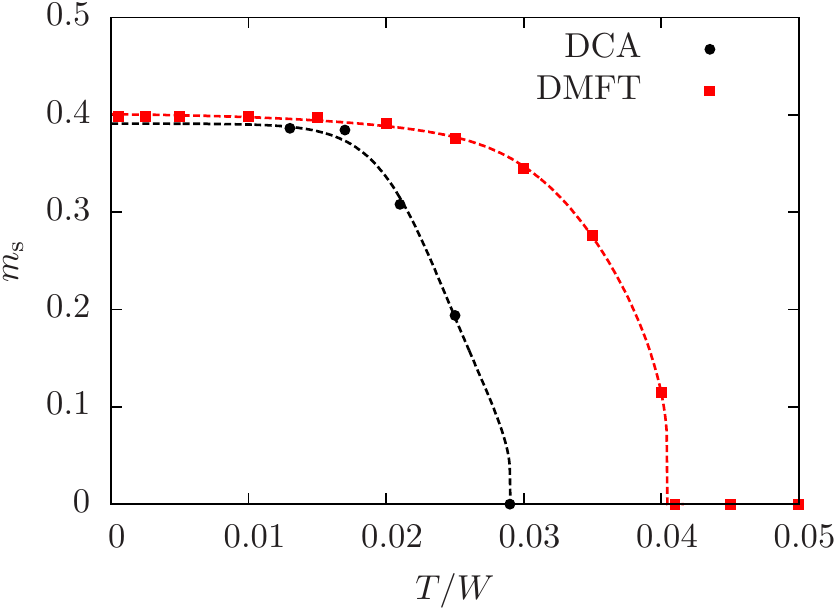}
  \caption{(Color online) Staggered magnetization $m_s(T)$ as function of $T$ for
    $U=0.67W$ as obtained from a DCA calculation with $N=18$ (circles)
    and the DMFT (squares). Dashed lines: guide to the
    eye. \label{fig:moft}}
\end{figure}

A first simple test of the method is to calculate the staggered moment
\begin{equation}
  m_s = \sum\limits_i m_i\,\exp \left( {\rmi {\bs Q}\cdot{\bs r}_i} \right)
\end{equation}
and thus locate the antiferromagnetic phase.  The results of
such a calculation for $U=0.67\,W$ as function of temperature are shown
in Fig.~\ref{fig:moft}
for DCA simulations (circles) and, for comparison, DMFT calculations
using Wilson's numerical renormalization group algorithm as the
impurity solver\cite{bullareview} (squares). The first thing to note
is that within DCA the critical temperature is reduced by roughly 30\%
as compared to the DMFT. The values of $T_N\approx 0.03\,W$ for DCA
and $T_N\approx0.042\,W$ for DMFT nicely agree with the results
obtained by Kent {\it et al.}\cite{kent}. This is in agreement with
the expectation, that for a three-dimensional system far enough away
from the critical region one should not see dramatic influence by the
order parameter fluctuations any more. Note, however, that
$m_s(0)\approx0.39$ for both DMFT and DCA is reduced as compared to
the Hartree approximation, where $m_söH(0)\approx0.426$.  Finally,
while the functional shape of $m_s(T)$ for the DMFT nicely follows the
standard mean-field behavior $m_s(T\nearrow T_N)\propto
\sqrt{1-T/T_N}$ and, respectively, $m_s(T\to0)\propto 1-2\rme^{-2T_N/T}$,
the form obtained from DCA is very different, rather exhibiting a
linear behavior just below $T_N$ and a constant value for $T\lesssim
0.02\,W$.

\begin{figure}
  \centering
  \includegraphics{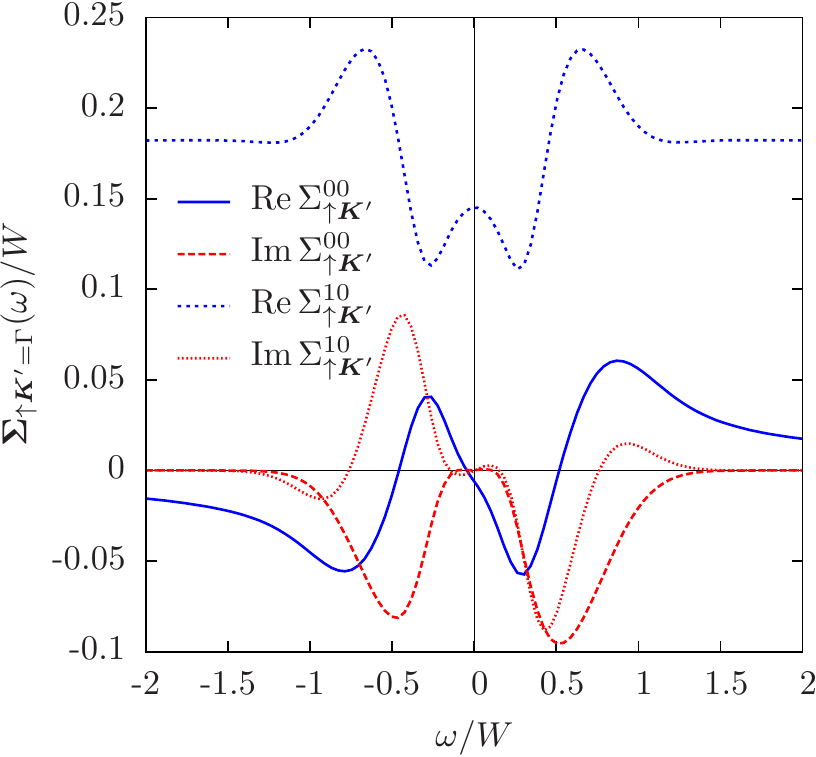}
  \caption{(Color online) Self-energy for ${\bs K}'=\Gamma$,
    $U=0.5\,W$, $t'=0$, and $T=0.02\,W$ in the antiferromagnetic
    phase. The real and imaginary parts of the elements
    $\Sigma^{00}_{\uparrow \bs K'}(\omega) = \Sigma^{11}_{\downarrow
      \bs K'}(\omega) = \Sigma^{11}_{\uparrow \bs K'}(-\omega) =
    \Sigma^{00}_{\downarrow \bs K'}(-\omega)$ and
    $\Sigma^{10}_{\uparrow \bs K'}(\omega) = \Sigma^{01}_{\uparrow \bs
      K'}(\omega) = \Sigma^{10}_{\downarrow \bs K'}(\omega) =
    \Sigma^{01}_{\downarrow \bs K'}(\omega)$ are shown. }
  \label{fig:selfanti}
\end{figure}

With the ability to perform reliable calculations in the
symmetry-broken phase, one is of course interested in extracting
dynamics from the simulations, preferably by analytically continuing
the single-particle self-energy.  To this end we again need the
high-frequency behavior of the self-energy, which can be obtained from
a high-frequency expansion (see the Appendix\ref{app:tails}) as
\begin{widetext}
  \begin{equation}
    {\bf \Sigma}_{\sigma\bs K'}(\rmi\omega_n) = \Or((\rmi\omega_n)^{-2})+
    U
    \begin{pmatrix}
      \langle n_{\bar\sigma} \rangle -\frac{1}{2} & \langle m_{\bar\sigma} \rangle \\
      \langle m_{\bar\sigma} \rangle & \langle n_{\bar\sigma} \rangle -\frac{1}{2} \\
    \end{pmatrix}
    + \frac{U^2}{\rmi\omega_n}
    \begin{pmatrix}
      \langle n_{\bar\sigma} \rangle 
      \left(1-\langle n_{\bar\sigma} \rangle\right) 
      + \langle m_{\bar\sigma} \rangle^2 &
      \langle m_{\bar\sigma} \rangle 
      \left(1-2\langle n_{\bar\sigma} \rangle\right) \\
      \langle m_{\bar\sigma} \rangle 
      \left(1-2\langle n_{\bar\sigma} \rangle\right) &
      \langle n_{\bar\sigma} \rangle 
      \left(1-\langle n_{\bar\sigma} \rangle\right) 
      + \langle m_{\bar\sigma} \rangle^2 
    \end{pmatrix}
    \label{eq:antihigh}
  \end{equation}
\end{widetext}
using the staggered spin polarization 
\begin{equation}
  \langle m_{\sigma} \rangle = 
  \sum\limits_i \rme^{\rmi {\bs Q}\cdot{\bs r}_i}
  \langle n_{\sigma i}-n_{\bar\sigma i} \rangle\;.
\end{equation}
The direct analytic continuation of non-diagonal self-energies---or
Green's functions---is not possible, since the non-diagonal spectral
function $-\frac{1}{\pi}\Im\, \Sigma^{10}_{\sigma \bs
  K'}(\omega)$ has both negative and positive values while the
standard MEM algorithm can only deal with non-negative spectral 
functions. In order to solve this problem, we employ the linear
transformation \cite{tomczak} (omitting spin and frequency
dependencies)
\begin{equation}
\label{eq:trans}
  \Sigma^{\pm}_{\bs K'}  =
  \frac{\Sigma^{00}_{\bs K'} + \Sigma^{11}_{\bs K'}}{2} \pm
  \Sigma^{10}_{\bs K'} 
\end{equation}
and determine $\Im\,\Sigma^{\pm}_{\sigma\bs K'}(\omega)$ along
with the diagonal elements $\Im\,\Sigma^{00}_{\sigma\bs
  K'}(\omega)$ and $\Im\,\Sigma^{11}_{\sigma\bs K'}(\omega)$
using the MEM. As in the paramagnetic case, the high-frequency
coefficients of Eq.~(\ref{eq:antihigh}) are used to normalize the
self-energies prior to the analytic continuation. Finally, the real
parts are calculated by a Kramers-Kronig relation analogous to
Eq.~(\ref{eq:kramers}). Since the transformation (\ref{eq:trans}) is linear,
it holds for the analytically continued functions as well and can thus
be solved for the non-diagonal element $\Sigma^{10}_{\sigma
  K'}(\omega)$. An example for a complete self-energy matrix on the
real frequency axis obtained by this procedure is shown in
Fig.~\ref{fig:selfanti}.
Note that the diagonal and off-diagonal elements have different
symmetry properties. For the particle-hole symmetric situation
presented here, the former obey the relation
$\Sigma^{\alpha\alpha}_{\sigma\bs
  K'}(\omega+i0^+)=\Sigma^{\alpha\alpha}_{\bar{\sigma}\bs
  K'}(-\omega+i0^+)$ and, respectively,
$\Sigma^{\bar{\alpha}\bar{\alpha}}_{\sigma\bs
  K'}(\omega+i0^+)=\Sigma^{\alpha\alpha}_{\sigma\bs
  K'}(-\omega+i0^+)$ following from the structure of the N\'eel
state, while the latter are all identical but obey
$\Sigma^{\alpha\bar{\alpha}}_{\sigma\bs
  K'}(\omega+i0^+)=-\Sigma^{\alpha\bar{\alpha}}_{\bar{\sigma}\bs
  K'}(-\omega+i0^+)$. This last relation in particular implies that
the real part is an even function of $\omega$, and the imaginary part
is odd.

The resulting self-energies for the cluster $\bs K'$ points are then
interpolated as in the paramagnetic case, and finally the
spin-averaged spectral function for all momenta $\bs k'$ of the
magnetic Brillouin zone follows from
\begin{multline}
  A_{\bs k'}(\omega) = -\frac{1}{\pi}\Im\,\Tr\,\\
  \left[ \left(
  \begin{array}{l}
    \omega+\mu-\epsilon_{\bs k'} \qquad\ \  0 \\
    0 \hfill \omega+\mu-\epsilon_{{\bs k}'+{\bs Q}} 
  \end{array}
  \right)
  - {\bf \Sigma}_{{\bs k}'}(\omega) \right]^{-1}\;,
\end{multline}
where Tr denotes the trace over the 2$\times$2-matrix.

\begin{figure*}
  \includegraphics{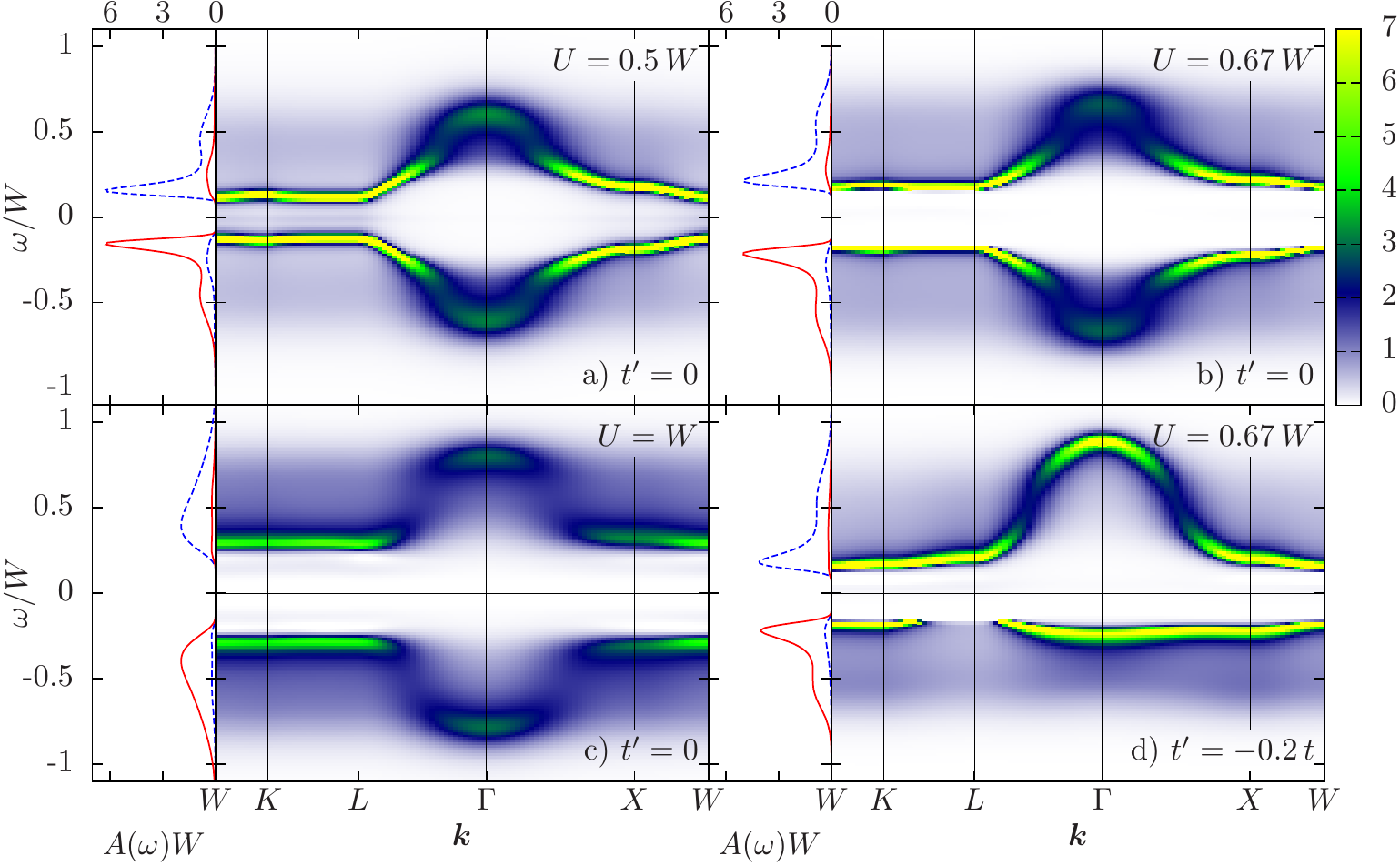}
  \caption{(Color online) Spin-averaged single-particle spectra for
    $T=0.02\,W$ in the antiferromagnetic phase for different
    interaction strengths. Panel (d) shows a spectrum
    for non-zero next-nearest-neighbor hopping $t'=-0.2\,t$. The left
    part of each figure depicts the local single-particle spectrum for
    both the majority spins (full red line) and the minority spins
    (dashed blue line). In (a) the edges of the gap are too sharp
    to resolve properly. To avoid a numerical division by
    zero, an artificial imaginary shift $-\rmi\delta$ with
    $\delta=0.03\,W$ was added to the self-energy. The result is a
    slight broadening of the gap edges. The interpolation follows the
    path along the high symmetry points of the reduced Brillouin zone
    depicted in Fig.~\ref{fig:selfanti}(b).}
  \label{fig:anti}
\end{figure*}

Results for single-particle spectra in the antiferromagnetically
ordered phase and different values of $U$ for $T=0.02\,W$ are shown in
Figs.~\ref{fig:anti}(a)--\ref{fig:anti}(c)
for paths connecting high symmetry points in the magnetic Brillouin
zone in Fig.~\ref{fig:selfanti}(b). As expected, the spectral function and
DOS have a gap around the Fermi energy, i.\,e., we always have an
insulating state. Furthermore, it is symmetric with respect to the
Fermi energy, reflecting the back-folding of the spectrum due to the
broken translational symmetry. Along a large part of the Brillouin
zone one has rather flat bands. For weak and moderate coupling these
structures have a rather high spectral weight, which results in the
formation of characteristic van Hove singularities at the gap
edges. This is a typical weak-coupling result consistent with a
conventional Hartree approximation. The gap increases for increasing
$U$, while at the same time the weight in the structures at the gap
edges is redistributed to larger energies, leading to a softening of
the structures in the DOS.

Figure~\ref{fig:anti}(d) shows an antiferromagnetic spectrum for non-zero
$t'=-0.2\,t$. The next-nearest-neighbor hopping breaks the symmetry
with respect to the Fermi energy analogous to the paramagnetic
case. Since magnetically frustrating interactions cause the
antiferromagnetic phase to quickly vanish at the present temperature,
no larger value of $t'$ was simulated.

\section{Conclusion}

The dynamical-cluster approximation is a controlled, systematically
improvable approach to the solution of strongly correlated electronic
model systems. Due to the size and complexity of the cluster impurity
problem, only quantum Monte Carlo methods are available as efficient
and unbiased quantum impurity solver algorithms.  While they yield in
principle arbitrarily accurate imaginary-frequency data, extracting
spectral functions from QMC data is an ill-posed numerical problem and
hence remains a difficult task.  It requires an analytical
continuation based on maximum entropy or a similar procedure and a
thorough error analysis for the quantity to be continued including
analytically supplemented high-frequency information. The previously
used Hirsch-Fye QMC impurity solver algorithm made the direct
continuation of the irreducible self-energy prohibitively expensive
and made comparatively unreliable root-searching techniques necessary.

The advent of modern continuous-time Monte-Carlo algorithms allows a
direct simulation of data in frequency space and moreover yields
high-quality data for the self-energy with reliable error estimates,
thus allowing a direct analytical continuation of the self-energy.

Based on this new route we presented a method to extract
momentum-resolved dynamical correlation functions from QMC simulations
of strongly correlated electron systems. We showed that QMC simulation
of clusters within the dynamical cluster approximation can provide
data accurate enough to enable the calculation of both momentum- and
frequency-resolved single-particle spectra. The method can resolve
detailed structures in the spectral functions, including
waterfall-like features.  We observe that even in three dimensions,
momentum resolved self-energies lead to spectra that are qualitatively
different from dynamical mean-field spectra, and present a more
reliable starting point for an extrapolation to the lattice system.

In addition we showed that we can access momentum- and
frequency-resolved spectra in the paramagnetic state as well as in
more complex ordered phases. As an example we discussed spectral
properties of the Hubbard model inside the antiferromagnetic phase,
also including an additional magnetic frustration introduced by a
next-nearest neighbor hopping.

We detected the interaction driven metal-insulator transition at
$T=0.012\,W_{\mathrm{eff}}$ and
$U_{\mathrm{c}}=0.94(3)\,W_{\mathrm{eff}}$, a value which is
substantially smaller than the DMFT result $U_{\mathrm{c}} =
1.18(2)\,W_{\mathrm{eff}}$. For non-zero values of $t'$ no MH-MIT
could be detected for temperatures $T\geq 0.01\,W$. The presence of a
smooth cross-over indicates that the transition has moved to lower
temperatures---an effect, that that has been previously studied within
the context of DMFT.\cite{peters_mit} To clarify this point as well as
the $t'$ dependence of $U_{\mathrm{c}}$ further studies at lower
temperatures are necessary. Furthermore, in contrast to the standard
expectation, the MH-MIT appears not to be purely local, but rather
occurs initially for $\bs k$ points at the Fermi surface only.  This
$\bs k$-dependent behavior is different from the momentum-selective
Mott transition found for the two-dimensional Hubbard
model\cite{selective} where the Mott transition occurs only on
selected parts of the Fermi surface. We do not observe such a behavior
in our data, but for a definite statement on this issue calculations
on larger clusters would be required.

\acknowledgements

  We thank Karlis Mikelsons for fruitful discussions. Our
  implementation of all algorithms is based on the libraries of the
  ALPS project \cite{alps2} and the ALPS DMFT project.\cite{ALPS_DMFT}
  ALPS (Applications and Libraries for Physics Simulations,
  \url{http://alps.comp-phys.org}) is an open source effort providing
  libraries and simulation codes for strongly correlated quantum
  mechanical systems.

  We acknowledge financial support by the Deutsche
  Forschungsgemeinschaft through the collaborative research center
  SFB~602 and by the German Academic Exchange Service (DAAD). This
  work was supported by the National Science Foundation through
  OISE-0952300, DMR-0706379, DMR-0705847, and through DMR-1006282. We
  used computational resources provided by the North-German
  Supercomputing Alliance (HLRN) and by the Gesellschaft f\"ur
  wissenschaftliche Datenverarbeitung G\"ottingen (GWDG).

\appendix*

\section{High-frequency expansion of the self-energies}
\label{app:tails}

We extend the results obtained for the dynamical mean field theory
\cite{tails1,tails2} to the momentum-dependent case encountered in the
DCA, both for the paramagnetic and the antiferromagnetic phase.

The antiferromagnetic coarse-grained Green's function $\bar{\bf
  G}_{\sigma \bs K'}(\rmi\omega_n)$ in the presence of a staggered
magnetic field can be described by \cite{clusterreview}
\begin{multline}
  \bar {\bf G}_{\sigma\bs K'}(\rmi\omega_n) 
  = \frac{1}{V} \int\!{\rmd}\tilde{\bs k} \\
  \left[
    \left(
      \begin{array}{l}
        \rmi\omega_n-\xi_{{\bs K'}+\tilde{\bs k}} \qquad\ \  h_{\sigma}/2 \\
        h_{\sigma}/2 \hfill \rmi\omega_n-\xi_{{\bs K}'+{\bs Q}+\tilde{\bs k}} 
      \end{array}
    \right)
    - {\bf \Sigma}_{\bs K'}(\rmi\omega_n)  \right]^{-1}
  \;
\end{multline}
using the matrix notation of Eq.~(\ref{eq:twobytwo}) and $\xi_{\bs K'} =
\epsilon_{\bs K'} -\mu$. To gain an expression for the
high-frequency coefficients of the self-energy, we use the ansatz
\begin{equation}
  {\bf \Sigma}_{\sigma \bf K'}(\rmi\omega_n) = 
  {\bf \Sigma}^0_{\sigma \bs K'} 
  + \frac{{\bf \Sigma}^1_{\sigma \bs K'}}{\rmi\omega_n}
  + \Or\left((\rmi\omega_n)^{-2} \right)
\;
\end{equation}
and expand the coarse-grained Green's function up to third order:
\begin{equation}
  \label{eq:greenexpand}
  \bar{\bf G}_{\sigma \bs K'}(\rmi\omega_n) = 
  \frac{{\bf C}^1_{\sigma \bs K'}}{ \rmi\omega_n } 
  +\frac{{\bf C}^2_{\sigma \bs K'}}{\left(\rmi\omega_n\right)^2} 
  +\frac{{\bf C}^3_{\sigma \bs K'}}{\left(\rmi\omega_n\right)^3} 
  + \Or\left((\rmi\omega_n)^{-4}\right)
\;.
\end{equation}
The result is
\begin{align}
  {\bf C}^1_{\sigma \bs K'} = & 
  \begin{pmatrix}
    1 & 0 \\ 0 & 1 
  \end{pmatrix}
  \;,
  \label{eq:kcoeff1}\\
  {\bf C}^2_{\sigma \bs K'} = & 
  \begin{pmatrix}
      \overline{\xi}_{\bs K'}
      &
      h_{\sigma}/2\\
      h_{\sigma}/2
      &
      \overline{\xi}_{{\bs K'}+{\bs Q}} \\
    \end{pmatrix}
  + {\bf \Sigma}^0_{\sigma\bs K'} \;,\\
  {\bf C}^3_{\sigma \bs K'} = & 
  \overline{
    \left[
      \begin{pmatrix}
        \xi_{\bs K'}
        &
        h_{\sigma}/2 \\
        h_{\sigma}/2
        &
        \xi_{{\bs K'}+{\bs Q}} \\
      \end{pmatrix}
      + {\bf \Sigma}^0_{\sigma\bs K'} 
    \right]^2 
  }+ {\bf \Sigma}^1_{\sigma\bs K'}
  \label{eq:kcoeff3}
  \;,
\end{align}
where the over-lined quantities are coarse grained over the momentum
patch centered around $\bs K'$, e.\,g.
\begin{equation}
  \overline{\xi}_{\bs K'} = \frac{1}{V} \int\!\rmd\tilde{\bs k}\,
  \xi_{{\bs K'} + \tilde{\bs k}}  \;.
\end{equation}

A direct calculation of the Green's function using Heisenberg's
equations of motion provides the information necessary for the
determination of the unknown coefficients ${\bf\Sigma}^0_{\sigma\bs
  K'}$ and ${\bf\Sigma}^1_{\sigma\bs K'}$. Starting again with the
Hamiltonian~(\ref{eq:hamfield}), the high frequency coefficients of the
single-particle Green's function in real space
\begin{equation}
  G_{\sigma ij}(\rmi\omega_n) = 
  \frac{C^1_{\sigma ij}}{ \rmi\omega_n } 
  +\frac{C^2_{\sigma ij}}{\left(\rmi\omega_n\right)^2} 
  +\frac{C^3_{\sigma ij}}{\left(\rmi\omega_n\right)^3} 
  + \Or\left((\rmi\omega_n)^{-4}\right)
\end{equation}
can be obtained \cite{tails2} via
\begin{align}
  C^1_{\sigma ij} = &
  \left\langle 
    \left\{
      c_{\sigma i}, c^{\dagger}_{\sigma j}
    \right\}
  \right\rangle\;, \\
  C^2_{\sigma ij} = &
  -\left\langle 
    \left\{ 
      \left[ H_h, c_{\sigma i} \right], c^{\dagger}_{\sigma j}
    \right\} 
  \right\rangle\;, \\
  C^3_{\sigma ij} = &
  \left\langle 
    \left\{ 
      \left[ 
        H_h, \left[ H_h, c_{\sigma i} \right]
      \right], c^{\dagger}_{\sigma j}
    \right\} 
  \right\rangle \;.
\end{align}
Here $[A,B]$ ($\{A,B\}$) denotes the (anti)commutator of the operators
$A$ and $B$. A straightforward calculation yields
\begin{align}
  \label{eq:coeff1}
  C^1_{\sigma ij} & =  
  \delta_{ij}\;,\\
  C^2_{\sigma ij} & =  
  -\tilde\xi_{ij} 
  -\left[
    h_{\sigma i} + U \langle n_{\bar\sigma i} \rangle
  \right] \delta_{ij} \;,\\
  C^3_{\sigma ij} & =  
  \sum\limits_{m} \tilde\xi_{im}\tilde\xi_{mj} + 
  \left( h_{\sigma i}h_{\sigma j} +
    U^2 \langle n_{\bar\sigma i} \rangle
  \right) \delta_{ij}
  \nonumber \\
  & - \left( h_{\sigma i} + h_{\sigma j} \right)\tilde\xi_{ij} 
  \nonumber\\
  &
  -U\left( n_{\bar\sigma i} + n_{\bar\sigma j} \right) 
  \left(\tilde\xi_{ij} - h_{\sigma i}\delta_{ij} \right)
  \;, \label{eq:coeff3}
\end{align}
where $\tilde\xi_{ij}=t+\left(\mu+\frac{U}{2}\right)\delta_{ij}$ for
nearest neighbors,
$\tilde\xi_{ij}=t'+\left(\mu+\frac{U}{2}\right)\delta_{ij}$ for
next-nearest neighbors, and zero otherwise. The high frequency
coefficients of the coarse-grained Green's function in cluster momentum
space [Eq.~(\ref{eq:greenexpand})] are readily calculated by a Fourier
transformation of Eqs. (\ref{eq:coeff1})--(\ref{eq:coeff3}) followed by a
coarse-graining in ${\bs k}$-space. One obtains
\begin{align}
  {\bf C}^1_{\sigma \bs K'} = & 
  \begin{pmatrix}
    1 & 0 \\ 0 & 1 
  \end{pmatrix}
  \\
  {\bf C}^2_{\sigma \bs K'} = & 
  \begin{pmatrix}
    \overline{\tilde\xi}_{\bs K'} + U\langle n_{\bar\sigma} \rangle  
    &
    U\langle m_{\bar\sigma} \rangle + h_{\sigma}/2\\
    U\langle m_{\bar\sigma} \rangle + h_{\sigma}/2
    &
    \overline{\tilde\xi}_{{\bs K'}+{\bs Q}} + U\langle
    n_{\bar\sigma} \rangle \\ 
  \end{pmatrix}
\end{align}
\begin{widetext}
  \begin{multline}
    {\bf C}^3_{\sigma \bs K'} =  
    \left(
      \begin{array}{l} 
        \overline{\tilde\xi^2}_{\bs K'} + h_{\sigma}^2/4 
        + 2U\langle n_{\bar\sigma} \rangle
        \overline{\tilde\xi}_{\bs K'}
        + U^2\langle n_{\bar\sigma} \rangle 
        + Uh_{\sigma}\langle m_{\bar\sigma} \rangle \\
        \left( U\langle m_{\bar\sigma} \rangle +h_{\sigma}/2 \right)
        \left( 
          \overline{\tilde\xi}_{\bs K'} 
          + \overline{\tilde\xi}_{{\bs K'} + {\bs Q}} 
        \right)
        + U^2\langle m_{\bar\sigma} \rangle 
        + Uh_{\sigma} \langle n_{\bar\sigma} \rangle
      \end{array}
    \right.\\
    \left.
      \begin{array}{r}
        \left( U\langle m_{\bar\sigma} \rangle +h_{\sigma}/2 \right)
        \left( 
          \overline{\tilde\xi}_{\bs K'} 
          + \overline{\tilde\xi}_{{\bs K'} + {\bs Q}} 
        \right)
        + U^2\langle m_{\bar\sigma} \rangle 
        + Uh_{\sigma} \langle n_{\bar\sigma} \rangle \\
        \overline{\tilde\xi^2}_{{\bs K'} + {\bs Q}} 
        + h_{\sigma}^2/4 
        + 2U\langle n_{\bar\sigma} \rangle
        \overline{\tilde\xi}_{{\bs K'} + {\bs Q}}
        + U^2\langle n_{\bar\sigma} \rangle 
        + Uh_{\sigma}\langle m_{\bar\sigma} \rangle      
      \end{array}
    \right)
    \;,
  \end{multline}
\end{widetext}
where $\tilde\xi_{\bs K'} = \xi_{\bs K'} - U/2$ and 
\begin{equation}
  \langle m_{\sigma} \rangle = 
  \sum\limits_i \rme^{\rmi {\bs Q}\cdot{\bs r}_i}
  \langle n_{\sigma i}-n_{\bar\sigma i} \rangle\;.
\end{equation} 
A comparison with
Eqs. \ref{eq:kcoeff1}--\ref{eq:kcoeff3} yields
\begin{equation}
  {\bf \Sigma}^0_{\bs K'}  = U
  \begin{pmatrix}
    \langle n_{\bar\sigma} \rangle -\frac{1}{2} & \langle m_{\bar\sigma} \rangle \\
    \langle m_{\bar\sigma} \rangle & \langle n_{\bar\sigma} \rangle -\frac{1}{2} \\
  \end{pmatrix}
\end{equation}
\begin{multline}
  {\bf \Sigma}^1_{\bs K'}  = U^2
  \left(
    \begin{array}{l}
      \langle n_{\bar\sigma} \rangle 
      \left(1-\langle n_{\bar\sigma} \rangle\right) 
      + \langle m_{\bar\sigma} \rangle^2 \\
      \langle m_{\bar\sigma} \rangle 
      \left(1-2\langle n_{\bar\sigma} \rangle\right) \\
    \end{array}
  \right.
  \\
  \left.
    \begin{array}{l}
      \langle m_{\bar\sigma} \rangle 
      \left(1-2\langle n_{\bar\sigma} \rangle\right) \\
      \langle n_{\bar\sigma} \rangle 
      \left(1-\langle n_{\bar\sigma} \rangle\right) 
      + \langle m_{\bar\sigma} \rangle^2 
    \end{array}
  \right)
  \;,
\end{multline}
which is the solution shown in Eq.~\ref{eq:antihigh}. The non-diagonal
parts vanish for $\langle m_{\sigma} \rangle=0$ and the expression
simplifies to the paramagnetic solution of Eqs. (\ref{eq:parahigh1})
and (\ref{eq:parahigh2}).

\bibliography{paper}

\end{document}